\documentclass[%
 aip,
 amsmath,amssymb,
reprint,%
groupedaddress,
]{revtex4-1}

\usepackage{url}
\usepackage{hyperref}

\usepackage{graphicx}
\usepackage{dcolumn}
\usepackage{bm}

\usepackage[utf8]{inputenc}
\usepackage[T1]{fontenc}
\usepackage{mathptmx}
\usepackage{etoolbox}

\newcommand{\be}{\boldsymbol{e}}

\newcommand{\bq}{\boldsymbol{q}}
\newcommand{\bn}{\boldsymbol{n}}
\newcommand{\bj}{\boldsymbol{j}}
\newcommand{\br}{\boldsymbol{r}}
\newcommand{\bx}{\boldsymbol{x}}

\newcommand{\bz}{\boldsymbol{z}}

\newcommand{\bB}{\boldsymbol{B}}

\newcommand{\bF}{\boldsymbol{F}}
\newcommand{\bG}{\boldsymbol{G}}

\newcommand{\bI}{\boldsymbol{I}}

\newcommand{\bL}{\boldsymbol{L}}

\newcommand{\bQ}{\boldsymbol{Q} }

\newcommand{\bU}{\boldsymbol{U}}

\newcommand{\polarorder}{\boldsymbol{m}}

\newcommand{\kt}{k_BT}

\newcommand{\Pe}{Pe}

\newcommand{\bUh}{\hat{\boldsymbol{U}} }

\usepackage{xcolor}

\usepackage[normalem]{ulem} 

\makeatletter
\def\@email#1#2{%
 \endgroup
 \patchcmd{\titleblock@produce}
  {\frontmatter@RRAPformat}
  {\frontmatter@RRAPformat{\produce@RRAP{*#1\href{mailto:#2}{#2}}}\frontmatter@RRAPformat}
  {}{}
}%
\makeatother
\begin{document}


\title[Forced microrheology of active colloids]{Forced microrheology of active colloids}
\author{Zhiwei Peng}
\author{John F. Brady}
\affiliation{ 
Division of Chemistry and Chemical Engineering, California Institute of Technology, Pasadena, California 91125, USA
}
\email{jfbrady@caltech.edu}

\date{\today}

\begin{abstract}
Particle-tracking microrheology of dilute active (self-propelled) colloidal suspensions is studied by considering the external force required to maintain the steady motion of an immersed constant-velocity colloidal probe. If the probe speed is zero, the suspension microstructure is isotropic but exhibits a boundary accumulation of active bath particles at contact due to their self-propulsion. As the probe moves through the suspension, the microstructure is distorted from the nonequilibrium isotropic state, which allows us to define a microviscosity for the suspension using the Stokes drag law. For a slow probe, we show that active suspensions exhibit a swim-thinning behavior in which their microviscosity is gradually lowered from that of  passive suspensions as the swim speed increases. When the probe speed is fast, the suspension activity is obscured by the rapid advection of the probe and the measured microviscosity is indistinguishable from that of passive suspensions. Generally for finite activity, the suspension exhibits a velocity-thinning behavior---though with a zero-velocity plateau lower than passive suspensions---as a function of the probe speed. These behaviors originate from the interplay between the suspension activity and the hard-sphere excluded-volume interaction between the probe and a bath particle. 
\end{abstract}

\maketitle

\section{Introduction}

In the past few decades, colloidal active matter systems such as motile bacteria and synthetic Janus particles immersed in a viscous solvent have evolved into a vibrant field of study \citep{Lauga_2009, Ramaswamy2010, Marchetti_review,elgeti2015physics,Bechinger2016rev}. Owing to their ability to self-propel, active colloids exhibit a cascade of striking properties not observed in equilibrium colloidal systems such as accumulation at no-flux boundaries \citep{Wensink2008, Li09, Elgeti_2013,yan_brady_2015}, upstream swimming in Poiseuille flow \citep{Hill07PRL, Koser09, Kaya12, Stark12,Peng2020}, and the existence of a steady-state spontaneous flow in the absence of any external forces \citep{Lushi2012,Guo2018}.

\begin{figure*}
  \centering
  \includegraphics[width=5in]{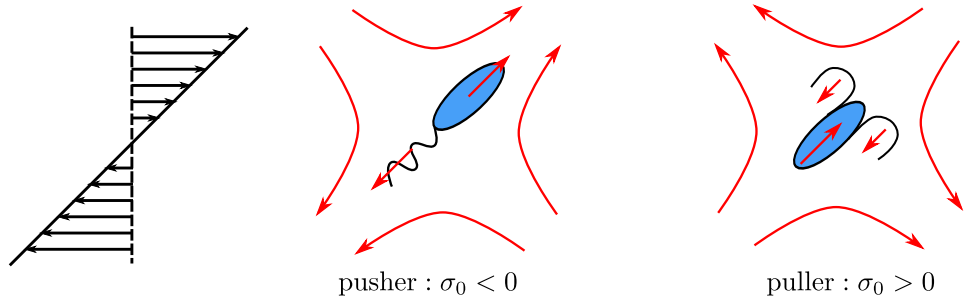}
  \caption{\label{fig:shear-schematic} Left: schematic of a simple shear flow. Center: schematic of the force dipole ($\sigma_0  < 0$) of a pusher  and the generated disturbance flow. Right: schematic of the force dipole ($\sigma_0  > 0$) of a puller  and the generated disturbance flow. For a pusher aligned with the extensional axis of the simple shear flow, the disturbance generated by its force dipole acts to stretch the fluid further. }
\end{figure*}

The macroscopic (bulk) rheological response of active colloidal suspensions is also distinct from that of passive colloids. In particular, experiments \citep{Lopez2015,Chui2021} and theoretical studies \citep{Hatwalne2004,Haines2009,saintillan2010,Ryan2011,Loisy2018} have shown that the low-$Pe$ shear (weak shear) viscosity of dilute active suspensions consisting of \emph{anisotropic} and \emph{pusher} (tail-actuated) microswimmers can be zero---or even negative. This apparent negative viscosity has been attributed to the interaction between the hydrodynamic stresslet induced by the force dipoles of an active particle and the applied simple shear flow. To see this, consider a self-propelled pusher or puller microswimmer in a simple shear flow (see Fig. \ref{fig:shear-schematic} for a schematic).  The straining-component of the flow tends to align the microswimmer along the extensional axis of shear. With this alignment, the flow induced by a pusher acts to ``stretch'' the fluid further, which results in a reduced shear viscosity. For a puller swimmer, the induced flow acts against the imposed shear and an increase in the shear viscosity is observed. According to Jeffery's equation describing the orientational dynamics of an ellipsoidal particle in simple shear \citep{jeffery}, this alignment effect is only present for nonspherical particles and, therefore, one expects the shear viscosity of active spheres to be nonnegative.

Bulk rheological studies such as simple shear provide a measurement of the global (suspension averaged) rheological behavior of colloidal suspensions. In the context of biological active matter such as cellular environments, the active ``particles'' are often subjected to spatially \emph{localized} cues and biochemical signals rather than to bulk flow or  body forces. These localized behaviors lead to an inherently heterogeneous intracellular environment with differing material properties such as spatial variations in viscosity and elasticity. In addition, classical bulk rheology equipment cannot be used to probe the microenvironment inside individual living cells without disrupting their mechanical structure. To address such challenges,  microrheological techniques have been developed \citep{furst2017microrheology}. In microrheology, the local rheological properties such as viscoelasticity of a complex fluid are inferred from the free (thermal) or forced motion of  ``probe'' particles. The probes can be either embedded colloidal particles or tagged organelles and molecules existing in the biological material. The study of the deformation or flow of biological materials at small length scales has been termed biomicrorheology and deemed a frontier in microrheology \citep{Weihs2006}. Indeed, particle-tracking microrheology has been widely used in experimental measurements to characterize the rheological properties inside living cells \citep{Wilhelm2003,Nawaz2012, berret2016local,ayala2016rheological,Hu2017}.

\begin{figure*}
\centering
\includegraphics[width=5in]{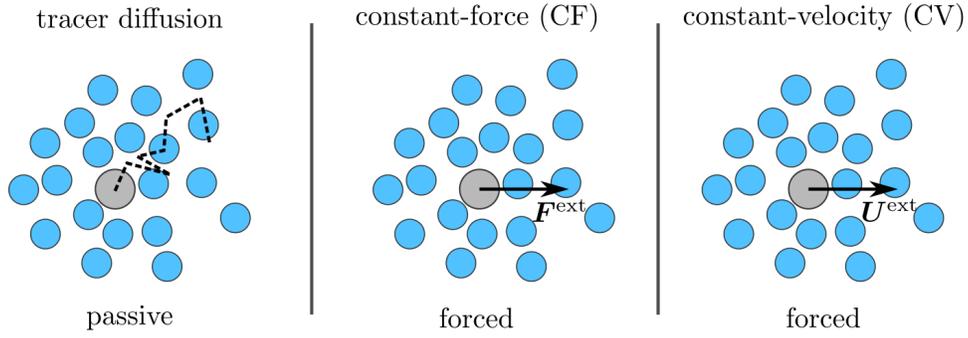}
\caption{\label{fig:micro-modes} Schematic of different operating modes of colloidal microrheology. The gray particle is the probe and the blue particles are the bath particles. In the passive mode, the mean squared displacement (MSD) of the probe is measured. In the forced modes, the probe is driven by an external force, either constant (CF mode) or fluctuating (CV mode).  }
\end{figure*}

To aid in the understanding of experimental measurements and in the prediction of colloidal microrheology, \citet{SquiresBrady2005} developed a theoretical framework in which a colloidal probe is pulled through a suspension of neutrally buoyant bath colloids (see Fig. \ref{fig:micro-modes} for a schematic). If the external pulling force is absent, this problem is often referred to as tracer diffusion and is classified as passive (no external forcing) microrheology. In the passive mode, the quantity of interest is often the long-time self-diffusivity of the probe (or tracer) particle. To characterize the nonlinear response, an external force, often larger than the thermodynamic restoring force, is applied to the probe and we call this problem forced microrheology \footnote{Traditionally, this is called \emph{active} microrheology in contrast to the passive mode. In the context of active matter, however, this terminology of active microrheology conflicts with that of active matter and we thus use the term forced microrheology.}. Within forced microrheology, two operating modes---constant-force (CF) and constant-velocity (CV)---are often considered from a theoretical perspective. In the CF mode, the external force $\bF^\text{ext}$ applied to the probe is a constant while the position of the probe is fluctuating. Conversely, for the CV mode, the velocity $\bU^\text{ext}$ is a constant (therefore, the position of the probe is known) and the external force required to maintain such a steady motion must fluctuate. The framework of  \citet{SquiresBrady2005} has been extended and used to study the microrheology of passive (not self-propelled) colloidal suspensions \citep{Khair2005viscoelastic,khair_brady_2006, meyer2006,zia_brady_2010,swan2013, Zia2018}.

In contrast to passive colloids, the study of the microrheology of active colloids is more recent and their microrheological response is less well understood \citep{Reichhardt2015, Burkholder2017,Burkholder2020,knevzevic2021oscillatory,Ahmed22}. Recent experimental and theoretical studies have shown that in the absence of external forcing, a Brownian tracer can undergo enhanced diffusive motion at long times due to  activity (e.g., self-propulsion) of the bath colloids, motile bacteria, or active enzymes. \citep{Jepson2013,mino2013induced,Morozov2014,Burkholder2017, Zhao2017}. It has also been shown that myosin motor protein activity can modify the rheological response of the filamentous actin \citep{LeGoff2001}. If the bath particles are passive, it is known that the  long-time diffusive motion of the tracer is hindered resulting from the ``collisions'' between the probe and the bath particles. This means that for active bath particles, activity-induced enhancement  is more than enough to overcome steric hindrance, which is present regardless of the activity of the bath particles. In the absence of hydrodynamic interactions (HIs),  \citet{Burkholder2017} showed that such enhanced diffusive motion can result from the interplay between the bath activity and the probe-bath steric interactions.

In forced microrheology, the viscous response of the suspension can be characterized by the effective microviscosity $\eta^\text{eff}$. Taking the CV mode as an example, one can relate the average external force to the probe velocity via $\left< \bF^\text{ext} \right>= 6 \pi \eta^\text{eff}a \bU_1$, where $a$ is the radius of the spherical probe and $\bU_1$ is the prescribed constant probe velocity. In the absence of HI, the microviscosity of passive colloids exhibit a velocity-thinning (or force-thinning in the CF mode) behavior as a function of the probe speed---the microscopic analog of shear-thinning \citep{SquiresBrady2005}. When the short-range hydrodynamic lubrication is considered, a force-thickening behavior can be observed  \citep{khair_brady_2006, swan2013} at large probe speed. For active Brownian suspensions without HI in the dilute limit, \citet{Burkholder2019JCP,Burkholder2020} studied forced microrheology by solving the Smoluchowski equation governing the distribution of active Brownian bath particles relative to the probe using a closure approximation. In particular, in the low probe speed limit a swim-thinning behavior is predicted \citep{Burkholder2019JCP}, and in the high probe speed limit, the microviscosity becomes indistinguishable from that of passive suspensions because the swimming motion is obscured by the much faster probe advection \citep{Burkholder2020}.

Following previous works \citep{Burkholder2017, Burkholder2019JCP, Burkholder2020}, we consider active colloidal suspensions modeled as monodisperse spherical active Brownian particles (ABPs) of radii $b$. The ABP model is one of the simplest descriptions for self-propelled particles. Furthermore, we study the CV mode of microrheology  in which the probe (particle $1$), which is a spherical colloidal particle of radius $a$,  has a prescribed CV $\bU_1$. In addition to the thermal Brownian motion with diffusivity $D_2$, the bath ABP (particle 2) self-propels with its intrinsic ``swim'' speed $U_2$ in a direction $\bq$, as illustrated in Fig. \ref{fig:cvmicro-schematic}. The orientation of swimming $\bq$ changes on a reorientation time scale $\tau_R$ that results from either continuous random Brownian rotations or the often-observed
discrete tumbling events of bacteria. The inverse of reorientation time defines a rotary diffusivity, $D_R=1/\tau_R$. One important intrinsic length scale due to activity is the run or
persistence length $\ell=U_2\tau_R$. Even for an isolated spherical ABP, the self-propulsion introduces a coupling between its orientational and translational dynamics, which is absent for a passive Brownian sphere. 

To characterize the microstructural deformation of the active suspension and the resulting viscous response, in the present work, we solve the \emph{full} Smoluchowski equation governing the probability distribution of a single ABP relative to the translating probe. This is done in the dilute limit in which only the interactions between the probe and one of the bath ABPs matter. Furthermore,  we neglect HIs between the probe and the ABP and focus on the interplay between the bath activity and the probe-ABP steric interaction. Resolving the full probability distribution allows us to examine the microstructure and the microviscosity in the full range of the ABP swim speeds and the probe speeds.

For a passive Brownian suspension, recall that the dimensionless microviscosity coefficient $\eta^\text{micro}$ [see Eqs. \eqref{eq:eta-micro-3d} and  \eqref{eq:eta-micro-2d}], which is proportional to the increment of viscosity $\eta^\mathrm{eff}-\eta$ normalized by the solvent viscosity $\eta$, exhibits a velocity-thinning behavior as a function of the probe P\'eclet number $Pe = U_1R_c/D_2$, where $R_c=a+b$ \citep{SquiresBrady2005}. The asymptotic results in the small and large $Pe$ limits are: $\eta^\text{micro}\to 1$ as $Pe \to 0$ and $\eta^\text{micro}\to 1/2$ as $Pe \to \infty$. For active suspensions, if the probe speed is much faster than the swim speed, the activity of the suspension is obscured by the probe advection and one recovers the passive result: $\eta^\text{micro}\to 1/2$ as $Pe \to \infty$. As such, the most interesting regime for the microrheology of active suspensions is small and intermediate $Pe$.

When the speed of the probe is small, the suspension is in the linear response regime (linear in terms of the probe speed, or $Pe$; see Sec. \ref{sec:slow-probe}) and the microviscosity obtained in this limit is called the zero-velocity microviscosity, $\eta_0^\text{micro}$ [see Eq.  \eqref{eq:CV-eta-0-limit}]. Corroborating the observations of \citet{Burkholder2019JCP}, we show that $\eta_0^\text{micro}$ decreases as the swim speed increases; therefore, the zero-velocity microviscosity exhibits a swim-thinning behavior. In the limit of no swimming, the passive result is recovered: $\eta_0^\text{micro} \to 1$ as $Pe_s=U_2R_c/D_2 \to 0$. When the swim speed is large, we show via a boundary layer analysis of the Smoluchowski equation that $\eta_0^\text{micro} \to 1/2$ as $Pe_s \to \infty$. Generally for finite activity, the microviscosity exhibits a velocity-thinning behavior as a function of $Pe$ but with a reduced $\eta_0^\text{micro}$ due to swim-thinning.

To obtain the results outlined thus far, in Sec. \ref{sec:CV-formulation} we formulate the problem from the Smoluchowski perspective. From the Smoluchowski equation governing the probability distribution of an ABP relative to the translating probe, we show that the microviscosity is related to the bath particle number density distribution at contact.  A perturbation analysis at small $Pe$ (Sec. \ref{sec:slow-probe}) allows us to derive equations governing the leading-order [$O(Pe)$] microstructure deformation, which gives rise to the zero-velocity microviscosity $\eta_0^\text{micro}$. In this small $Pe$ limit, we further consider the fast-swimming limit ($Pe_s \gg 1$) and show that $\eta_0^\text{micro}$ exhibits a swim-thinning behavior. In Sec. \ref{sec:cv-numerics}, we discuss the numerical methods used to solve the full Smoluchowski equation. To corroborate our analysis and numerical results from the Smoluchowski equation, in Sec. \ref{sec:CV-BD}, we consider Brownian dynamics (BD) simulations. In Sec. \ref{sec:cv-eta-micro}, we present the microviscosity as a function of the probe speed, the swim speed, and the reorientation time. We show that the results obtained by solving the Smoluchowski equation agree well with those from BD. We conclude in Sec. \ref{sec:cv-conclusion}. While our formulation applies in both two and three dimensions, the asymptotic analyses and numerical solutions are performed in two dimensions. This reduction in dimensionality allows us to solve the full Smoluchowski equation numerically. We note that the dimensionality does not affect the qualitative behavior of the measured microviscosity \citep{burkholder2019single}.

\section{Problem formulation}
\label{sec:CV-formulation}
We consider a dilute suspension of ABPs where only the pairwise interactions between the probe and a single ABP matter. At this level, the suspension microstructure in the presence of a probe is described by the pair probability distribution $P_2(\bx_1, \bx_2, \bq,t)$, where the positions of the probe ($\bx_1$) and the ABP ($\bx_2$) are in the laboratory frame. Because the probe has prescribed kinematics, i.e., CV, the position of the probe does not matter and the system is statistically homogeneous \citep{SquiresBrady2005}. Conditioning the pair probability on the position of the probe, we then have 
\begin{align}
    P_2(\bx_1, \bx_2, \bq, t) &= P_{1/1}(\br, \bq, t| \bz, t) P_1(\bz, t)\nonumber \\
    & = P_{1/1}(\br, \bq,t)P_1(\bz,t),
\end{align}
where $\br = \bx_2-\bx_1$, $\bz=\bx_1$. In other words, the conditional probability density function $P_{1/1}$ does not depend on the position of the probe in the laboratory frame.  As a result, it is most convenient to consider the conditional probability distribution of the ABP in a comoving frame that is attached to the probe particle. In this relative frame, the Smoluchowski equation governing the conservation of ABPs is written as \citep{Burkholder2019JCP, Burkholder2020}
\begin{align}
    \frac{\partial P_{1/1}(\br, \bq,t) }{\partial t} + \nabla_r \cdot\left(\bj_2^T - \bj_1^T\right) + \nabla_R \cdot\bj_2^R=0.
    \label{eq:CV-smol-conditional}
\end{align}
Here, $\nabla_r$ denotes the gradient operator in physical space ($\br$) and $\nabla_R = \bq\times \frac{\partial}{\partial \bq}$ is the gradient operator in orientation space.

\begin{figure*}
\centering
\includegraphics[scale=1.0]{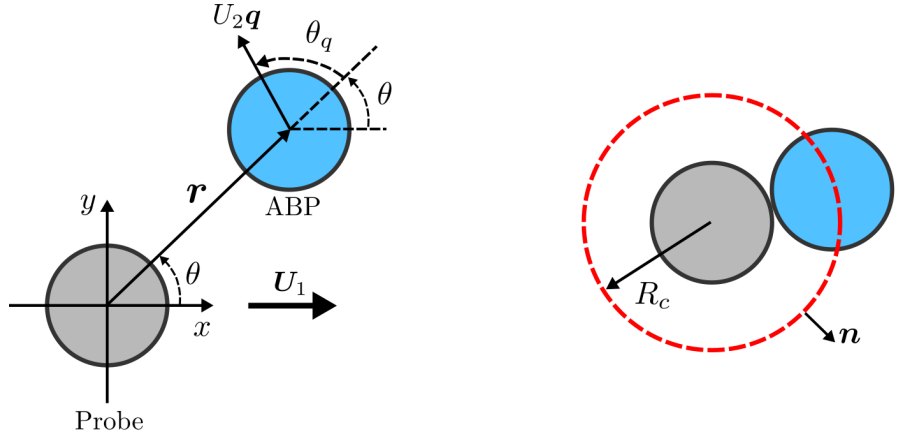}
\caption{\label{fig:cvmicro-schematic} Schematic of a probe particle driven at a CV $\bU_1$ and an active Brownian bath particle in 2D. The ABP swims with a constant speed $U_2$ in a direction $\bq$. The radius of contact is $R_c$ and $\bn$ is the unit normal vector pointing from the probe into the ABP suspension.}
\end{figure*}

In the absence of HIs between the probe and the ABP, the translational and rotational fluxes in the Smoluchowski Eq. \eqref{eq:CV-smol-conditional}, respectively, are 
\begin{align}
    \label{eq:CV-jT-P}
    \bj_2^T-\bj_1^T &= \left(U_2\bq - \bU_1\right) P_{1/1} -D_2 \nabla_r P_{1/1},\\
    \label{eq:CV-jR-P}
    \bj_2^R &= -D_R \nabla_R P_{1/1}.
\end{align}
Note that it is the flux of ABPs relative to the probe that appears in the equation so as to obey Galilean invariance. In the CV mode of microrheology, the relative diffusivity is simply the diffusivity of the ABP ($D_2$) instead of the sum of diffusivities of the probe and the ABP as in the CF mode of microrheology \citep{SquiresBrady2005}. Because we neglect HIs, the probe and ABP interact sterically due to their hard-sphere nature. That is, at the surface of contact ($|\br| =R_c= a+b$), the relative translational flux of particle centers vanishes:
\begin{align}
\label{eq:CV-no-flux-P}
    \bn \cdot\left(\bj_2^T-\bj_1^T\right) = 0,
\end{align}
where $\bn$ is the unit normal vector as shown in Fig. \ref{fig:cvmicro-schematic}.

Far from the probe, the suspension microstructure is undisturbed, giving 
\begin{equation}
\label{eq:CV-far-field-P}
    P_{1/1} \to \frac{n^\infty}{\Omega_d},
\end{equation}
where $n^\infty$ is the number density in the undisturbed ABP suspension, which has the units of number per volume (or area in 2D), and $\Omega_d$ is the total solid angle of the orientation space in $d$ dimensions. In 2D, the solid angle $\Omega_2 = 2\pi$; in 3D, $\Omega_3 = 4\pi$.

Equation \eqref{eq:CV-smol-conditional} together with its flux expressions \eqref{eq:CV-jT-P}--\eqref{eq:CV-jR-P} and boundary conditions  \eqref{eq:CV-no-flux-P}--\eqref{eq:CV-far-field-P} governs the disturbance of the suspension microstructure due to the CV probe. In the absence of activity, by setting $U_2=0$ and integrating out the orientational degree of freedom, one can recover the CV microrheology problem for passive Brownian suspensions considered by  \citet{SquiresBrady2005}.

In the CV mode of microrheology, the main quantity of interest is the external force required to maintain such a steady probe motion. Due to the fluctuating nature of the ABP suspension, the external force averaged over Brownian fluctuations is considered. For an active Brownian suspension, this has been shown to be  \citep{Burkholder2019JCP}
\begin{align}
\label{eq:F-ext-P}
    \left< F^\text{ext}\right> = \zeta_1U_1 + \kt \oint _{r=R_c} \bn\cdot\bUh_1P_{1/1} dS d\bq,
\end{align}
where $\zeta_1$ is the Stokes drag of the probe, $\kt$ is the thermal energy, and $\bUh_1$ is a unit vector in the direction of the probe motion, $\bUh_1 = \bU_1/|\bU_1|$. The only difference between Eq. \eqref{eq:F-ext-P} and that obtained for a passive Brownian suspension \citep{SquiresBrady2005} is the additional integration over the orientational degrees of freedom of the active bath particle. In Eq. \eqref{eq:F-ext-P}, only the net force in the direction of the probe motion is considered because the net force in the transverse direction vanishes due to symmetry. The applied external force in a hard-sphere passive colloidal suspension is larger than the Stokes drag ($\zeta_1 U_1$) of the probe due to the fact that in order to maintain a CV,  the probe has to push away bath particles along its trajectory.  To quantify this increase in the applied force, it is convenient to consider the dimensionless viscosity increment defined as 
\begin{align}
\label{eq:CV-viscosity-increment}
    \frac{\Delta \eta }{\eta} =  \frac{\left< F^\text{ext}\right> - \zeta_1 U_1} {\zeta_1 U_1},
\end{align}
where an effective microscopic viscosity can be defined via $\left< F^\text{ext}\right> = 6 \pi \eta^\text{eff} a U_1$ and equivalently $\Delta \eta /\eta = (\eta^\text{eff} - \eta )/\eta$ with $\eta$ being the viscosity of the solvent. By definition, the viscosity increment vanishes in the absence of bath particles.

A dimensional analysis reveals four time scales that govern the microrheology of active Brownian suspensions: (1) the diffusive time scale $\tau_D = R_c^2/D_2$, (2) the swim time scale $\tau_S = R_c/U_2$, (3) the advective time scale $\tau_\text{adv} = R_c/U_1$, and (4) the reorientation time $\tau_R=1/D_R$. Comparing the other three time scales with the diffusive time scale gives three dimensionless groups. The first one is the swim P\'eclet number given by 
\begin{equation}
    Pe_s = \frac{\tau_D}{\tau_S} = \frac{U_2R_c}{D_2}.
\end{equation}
The second dimensionless group is the  P\'eclet number of the probe (using the diffusivity of the ABP) 
\begin{equation}
    Pe = \frac{\tau_D}{\tau_\text{adv} }= \frac{U_1R_c}{D_2}.
\end{equation}
Finally, comparing $\tau_R$ with $\tau_D$ defines the third parameter,
\begin{equation}
    \gamma =\left( \frac{\tau_D}{\tau_R}\right)^{1/2} = \frac{R_c}{\delta},
\end{equation}
where $\delta = \sqrt{D_2\tau_R}$ is the microscopic diffusive step taken by the ABP on the reorientation time scale $\tau_R$.

To render the equations dimensionless, we scale lengths by $R_c$ and time by $\tau_D$ and define the dimensionless probability distribution (or the suspension microstructure) $g$ such that 
\begin{align}
    P_{1/1}(\br, \bq,t) = n^\infty g(\br, \bq, t). 
\end{align}
From Eqs. \eqref{eq:CV-smol-conditional}, \eqref{eq:CV-jT-P}, and \eqref{eq:CV-jR-P}, we obtain the Smoluchowski equation  for the microstructure as 
\begin{align}
\label{eq:CV-smol-g}
    \frac{\partial g}{\partial t} + \nabla_r\cdot\left[ \left(Pe_s \bq - Pe \bUh_1\right) g - \nabla_r g\right] - \gamma^2 \nabla_R^2 g=0. 
\end{align}
The no-flux condition \eqref{eq:CV-no-flux-P} becomes 
\begin{equation}
\label{eq:CV-no-flux-g}
    \bn \cdot\left[ \left(Pe_s \bq - Pe \bUh_1\right) g - \nabla_r g\right]=0\quad\text{at}\quad r=1,
\end{equation}
and the far-field condition \eqref{eq:CV-far-field-P} translates into 
\begin{equation}
\label{eq:CV-far-field-g}
    g \to \frac{1}{\Omega_d} \quad\text{as}\quad r \to \infty.
\end{equation}

Making use of the Stokes-Einstein-Sutherland relation $k_BT = D_2\zeta_2 = D_2 6 \pi \eta b$ and the definition of volume fraction $\phi = 4\pi b^3 n^\infty/3$ in 3D, we obtain from \eqref{eq:F-ext-P} and \eqref{eq:CV-viscosity-increment} the scaled viscosity increment
\begin{align}
    \frac{\Delta \eta}{\phi\eta} = \frac{3}{4 \pi}\frac{D_2 R_c^2}{ab^2 U_1}\oint_{r=1} \bn\cdot\bUh_1 ~n~ d S ,
\end{align}
where 
\begin{equation}
    n(\br, t) = \int g(\br, \bq,t) d \bq
\end{equation}
is the dimensionless number density, which tends to unity as $r\to \infty$. Noting that
\begin{align}
    \frac{D_2 R_c^2}{ab^2 U_1}  = \frac{(1+\alpha)^3}{\alpha Pe},
\end{align}
and $\alpha = a/b$, we obtain
 \begin{align}
 \label{eq:CV-scaled-viscosity-increment}
     \frac{\Delta \eta}{\phi\eta}  = \frac{3}{4\pi}  \frac{(1+\alpha)^3}{\alpha Pe}\oint_{r=1} \bn\cdot\bUh~n~d S.
 \end{align}
Noting that the microstructure only depends on the contact radius $R_c$ instead of the sizes of both the probe and the ABP [see Eq. \eqref{eq:CV-smol-g}], the only dependence of the scaled viscosity increment $\Delta\eta/(\phi\eta)$ on the size ratio $\alpha$ is in the prefactor before the integral in Eq. \eqref{eq:CV-scaled-viscosity-increment}. Therefore, for convenience we define the so-called effective microviscosity coefficient as 
\begin{equation}
\label{eq:eta-micro-3d}
    \eta^\text{micro} = \frac{\Delta \eta }{\phi \eta} \frac{2 \alpha }{(1+\alpha)^3} = \frac{3}{2\pi Pe}\bUh_1\cdot\oint_{r=1}\bn~n~dS.
\end{equation}
Here, a factor of $2$ is introduced in front of the factor $\alpha/(1+\alpha)^3$ so that for a passive Brownian suspension $\eta^\text{micro} \to 1$ as $Pe \to 0$ \citep{SquiresBrady2005}. By construction, $\eta^\text{micro}$ does not depend on the size ratio $\alpha$. In 2D, we use the area fraction $\phi =n^\infty  \pi b^2 $ and obtain a similar definition
\begin{equation}
\label{eq:eta-micro-2d}
       \eta^\text{micro}_{2D} = \frac{\Delta \eta }{\phi \eta} \frac{ \alpha }{(1+\alpha)^2} = \frac{1}{\pi Pe}\bUh_1\cdot\oint_{r=1}\bn~n~dS. 
\end{equation}
Hereinafter, we use microviscosity and microviscosity coefficient interchangeably to refer to the effective microviscosity coefficient defined above.

Because the bath particles are active, the phase space of the microstructure includes both the relative position $\br$ and the orientation $\bq$. The high dimensionality of the phase space is challenging for the numerical  simulation of the Smoluchowski Eq. \eqref{eq:CV-smol-g}. In 3D, by parametrizing the orientation vector $\bq$ using the polar and azimuthal angles of a spherical coordinate system, the phase space has a dimensionality of $5$: three dimensional in space and two dimensional in orientation.  As shown in previous work \citep{yan_brady_2015, Burkholder2019JCP}, the dimensionality only affects the solution of the Smoluchowski equation in a quantitative manner. In this paper, we focus on the microrheology of ABPs in 2D.

\subsection{The Smoluchowski equation in 2D}
\label{sec:CV-2D-equations}
Equation \eqref{eq:CV-smol-g} is most convenient for analysis in a polar coordinate system for the physical space and in a relative angular coordinate system for the orientation space in which $\bq = \cos\theta_q \be_r + \sin\theta_q \be_\theta$ (see Fig. \ref{fig:cvmicro-schematic} for a schematic). Here, $\be_r = \cos\theta \be_x + \sin\theta\be_y$  is the radial basis vector in the polar coordinate system and $\be_\theta$ is the basis vector in the angular ($\theta$) direction. Without loss of generality, we take $\bUh_1=\be_x$ so that the probe moves in the positive $x$ direction. In this $(r, \theta, \theta_q)$ coordinate system, Eq. \eqref{eq:CV-smol-g} is written explicitly as 
\begin{align}
\label{eq:CV-smol-g-2D-explicit-eq}
    &\left( Pe_s  \cos\theta_q - Pe \cos\theta\right) \frac{\partial g}{ \partial r} \nonumber \\
    &+\left( Pe_s\sin\theta_q +Pe\sin\theta\right) \frac{1}{r}\left(\frac{\partial g}{\partial \theta} - \frac{\partial g}{\partial \theta_q} \right)\nonumber\\
    &- \frac{1}{r}\frac{\partial}{\partial r} r \frac{\partial g}{\partial r} - \frac{1}{r^2}\left( \frac{\partial ^2 g}{\partial \theta^2}-2\frac{\partial^2 g}{\partial \theta \partial \theta_q} +\frac{\partial ^2 g}{\partial \theta_q^2} \right)\nonumber \\
    & -\gamma^2 \frac{\partial^2 g}{\partial \theta_q^2} =0,\\
    \label{eq:no-flux-g-2d-explicit}
    & \left( Pe_s  \cos\theta_q - Pe \cos\theta\right)g - \frac{\partial g}{\partial r} =0 \quad \text{at}\quad r=1,\\
    \label{eq:g-2d-explicit-far}
    & g \to \frac{1}{2\pi}\quad\text{as}\quad  r \to \infty.
\end{align}

\section{A slow probe}
\label{sec:slow-probe}
\subsection{Perturbation expansion of the microstructure}
For a slow probe ($Pe \ll 1$), the suspension microstructure is only slightly displaced from the state in which the probe is held fixed in a bath of ABPs. This allows us to consider the microstructure in the perturbation series $g = g_0(\br, \bq) +Pe g_1(\br, \bq)+\cdot\cdot\cdot$.  The governing equations at $O(1)$ and $O(Pe)$ in any dimension are given by Eqs. \eqref{eq:CV-g0-operator-form}--\eqref{eq:CV-g1-bc-operator-form} in appendix \ref{sec:CV-moments-appendix}.

In 2D, Eqs. \eqref{eq:CV-g0-operator-form}--\eqref{eq:CV-g1-bc-operator-form} can be written explicitly in the $(r, \theta, \theta_q)$ frame. To this end, we first write the perturbation series as $g = g_0(r, \theta_q) +Pe g_1(r, \theta, \theta_q)+\cdot\cdot\cdot$. At $O(1)$, the probe is held stationary (zero-velocity) and the problem reduces to that of ABPs in the exterior of a disk \citep{yan_brady_2015}, which exhibits spherical symmetry and thus $g_0$ is independent of the angular position $\theta$. The probability distribution of ABPs outside of a fixed probe is governed by 
\begin{align}
    \label{eq:CV-slow-probe-g0}
   & Pe_s  \cos\theta_q \frac{\partial g_0}{ \partial r} -Pe_s\sin\theta_q \frac{1}{r}\frac{\partial g_0}{\partial \theta_q} - \frac{1}{r}\frac{\partial}{\partial r} r \frac{\partial g_0}{\partial r} \nonumber \\ 
   &- \left( \frac{1}{r^2} +\gamma^2\right)\frac{\partial ^2 g_0}{\partial \theta_q^2}   =0,\\
   & Pe_s \cos\theta_q g_0 - \frac{\partial g_0}{\partial r} =0 \quad \text{at}\quad r=1, \\
   &  g_0 \to \frac{1}{2\pi}\quad \text{as}\quad r \to \infty.
\end{align}

The first effect of the probe motion appears at $O(Pe)$, which is  governed by  
\begin{align}
\label{eq:CV-slow-probe-f1-eq}
   &Pe_s  \cos\theta_q \frac{\partial f_1}{ \partial r} +Pe_s\sin\theta_q \frac{1}{r}\left(f_2-\frac{\partial f_1}{\partial \theta_q} \right)\nonumber\\
   &- \frac{1}{r}\frac{\partial}{\partial r} r \frac{\partial f_1}{\partial r} -  \frac{1}{r^2} \left(-f_1 -2 \frac{\partial f_2}{\partial \theta_q}  +\frac{\partial^2 f_1}{\partial \theta_q^2}\right)\nonumber\\
   &-\gamma^2\frac{\partial ^2 f_1}{\partial \theta_q^2}   =\frac{\partial g_0}{\partial r},\\
   \label{eq:CV-slow-probe-f2-eq}
   &Pe_s  \cos\theta_q \frac{\partial f_2}{ \partial r} +Pe_s\sin\theta_q \frac{1}{r}\left(-f_1-\frac{\partial f_2}{\partial \theta_q} \right)\nonumber\\
   &- \frac{1}{r}\frac{\partial}{\partial r} r \frac{\partial f_2}{\partial r} -  \frac{1}{r^2} \left(-f_2 +2 \frac{\partial f_1}{\partial \theta_q}  +\frac{\partial^2 f_2}{\partial \theta_q^2}\right)\nonumber\\
   &-\gamma^2\frac{\partial ^2 f_2}{\partial \theta_q^2}   =\frac{1}{r}\frac{\partial g_0}{\partial \theta_q},
\end{align}
where $g_1(r, \theta, \theta_q) = f_1(r, \theta_q)\cos\theta + f_2(r, \theta_q)\sin\theta$. The no-flux condition reduces to
\begin{align}
    \label{eq:Pe-1-bc-f1}
Pe_s \cos\theta_q f_1 - \frac{\partial f_1}{\partial r} &= g_0\quad \mathrm{at}\quad r=1.\\
\label{eq:Pe-1-bc-f2}
 \quad Pe_s\cos\theta_q f_2 - \frac{\partial f_2}{\partial r}&=0\quad \mathrm{at}\quad r=1.
\end{align}
Using Eq. \eqref{eq:eta-micro-2d}, we obtain 
\begin{align}
\label{eq:CV-eta-0}
    \eta_0^\text{micro} = \int_0^{2\pi} f_1(r=1, \theta_q)d\theta_q,
\end{align}
where 
\begin{equation}
\label{eq:CV-eta-0-limit}
    \eta_0^\text{micro} = \lim_{Pe \to 0} \eta^\text{micro}
\end{equation}
is the microviscosity in the limit $Pe \to 0$, or the zero-velocity microviscosity. The $O(Pe)$ number density, $n_1(\br) =\int g_1d\bq$, has the form $n_1(\br) = \bUh_1\cdot\br p_3(r)$ due to symmetry, where $p_3$ is an unknown scalar function of $r$ (see appendix \ref{sec:CV-moments-appendix}). From this and \eqref{eq:eta-micro-2d}, we see that $\eta_0^\text{micro} = p_3(1)$ in 2D.

In general, $\eta_0^\text{micro}$ for active Brownian suspensions is a function of $Pe_s$ and $\gamma$. For passive suspensions ($Pe_s=0$), the orientational distribution is uniform and the density at $O(Pe)$ is $n_1 = \bUh_1\cdot\be_r /r$ in 2D, which gives $\eta_0^\text{micro}=1$.
If the suspension is weakly active, we expect the zero-velocity microviscosity to approach that of a passive suspension. That is,  $\eta_0^\text{micro} \to 1$ as $Pe_s \to 0$.

\subsection{Fast-swimming ABPs}
\label{sec:fast-swimming}
We now consider the suspension microstructure and the zero-velocity microviscosity in the fast-swimming limit characterized by $Pe_s\gg 1$. The rotational diffusivity is assumed to be finite, $\gamma \sim O(1)$. In this high $Pe_s$ limit, translational diffusion is only important in a boundary layer near the probe. The boundary layer thickness is dictated by the balance between the swimming flux into the probe and the diffusive flux down the concentration gradient, which gives a thickness of $O(Pe^{-1}_s R_c)$. Therefore, we define a stretched boundary layer coordinate $\rho = (r-1)/\epsilon$, $\epsilon = 1/Pe_s$, such that $\rho \sim O(1)$ as $r \to 1$ and $\epsilon \to 0$.

In the boundary layer, the governing equation for $g_0(\rho, \theta, \theta_q)$ becomes
\begin{align}
  &\cos\theta_q \frac{\partial g_0 }{\partial \rho} -  \sin\theta_q\frac{\epsilon }{1+\rho \epsilon} \frac{\partial g_0}{\partial \theta_q} - \frac{\epsilon}{1+\rho\epsilon}\frac{\partial g_0}{\partial \rho}\nonumber \\
  &- \frac{\partial^2 g_0}{\partial \rho^2} - \epsilon^2 \left( \gamma^2 + \frac{1}{(1+\rho\epsilon)^2}\right) \frac{\partial^2 g_0}{\partial \theta_q^2}=0,\label{eq:fix_probe_g0}\\
  & \cos\theta_q g_0 - \frac{\partial g_0}{\partial \rho }=0 \quad \text{at}\quad \rho=0,\\
  & g_0 \to \frac{1}{2\pi} \quad \text{as}\quad \rho \to \infty.
\end{align}
To study the microstructure for a stationary probe in the fast-swimming limit, we pose the perturbation expansion $g_0 = \epsilon^{-1} g_0^{(-1)} + g_0^{(0)} + \epsilon g_0^{(1)}+o(\epsilon)$, where the leading-order microstructure is  $O(1/\epsilon)$ (singular) as $\epsilon \to 0$ \citep{Yan2018,peng2021activity}. Inserting this expansion into equation (\ref{eq:fix_probe_g0}) yields equations for $g_0^{(k)}$ ($k=-1,0,...$). 

The solution to $g_0^{(-1)}$ can be readily obtained as 
\begin{align}
\label{eq:CV-g0m1}
  g_0^{(-1)} = A_1(\theta_q)e^{\rho\cos\theta_q}.
\end{align}
Here, $A_1$ is an unknown function of $\theta_q$ that will be determined from the solution at the next order. Because the distribution is $O(1)$ far from the probe, we require $g_0^{(-1)} \to 0$  as $\rho \to \infty$. This means that the solution is only valid in the region $\cos\theta_q<0$. Outside this region in the orientation space, the solution is zero at this order. Physically, this is due to the fact that ABPs in contact with the probe have to point toward the probe, i.e. $\bq\cdot\be_r = \cos\theta_q <0$,  because otherwise they would swim away.

\begin{figure*}
  \centering
  \includegraphics[width=5.5in]{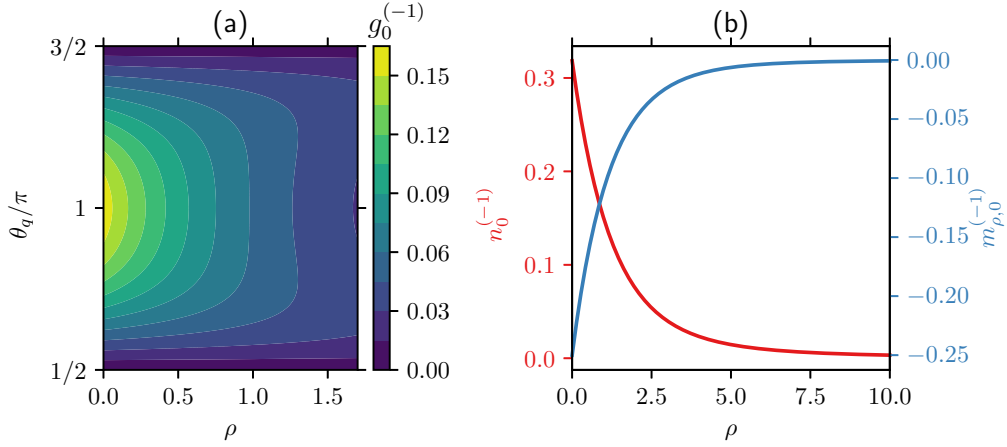}
  \caption{\label{fig:g0m1} (a) Contour plot of the leading-order distribution function $g_0^{(-1)}$ as a function of $\rho$ and $\theta_q$. (b) The leading-order number density $n_0^{(-1)}$ and polar order in the radial direction $m_{\rho,0}^{(-1)}$ as a function of $\rho$.}
\end{figure*}

At $O(1)$, we have 
\begin{align}
\label{eq:CV-slow-probe-g00-eq}
  &\cos\theta_q \frac{\partial g_0^{(0)} }{\partial \rho} - \frac{\partial^2 g_0^{(0)}}{\partial \rho^2} = \sin\theta_q\frac{\partial g_0^{(-1)}}{\partial \theta_q} + \frac{\partial g_0^{(-1)}}{\partial \rho},\\
  & \cos\theta_q g_0^{(0)} - \frac{\partial g_0^{(0)}}{\partial \rho} =0 \quad \text{at}\quad \rho =0,\\
  & g_0^{(0)} \to \frac{1}{2\pi}\quad \text{as} \quad \rho \to \infty.
\end{align}
The far-field condition on $g_0^{(0)}$ ensures proper matching with the constant solution outside the boundary layer. The general solution can be written as 
\begin{align}
\label{eq:CV-slow-probe-g00}
g_0^{(0)} = &\left(B_1(\theta_q) \rho +B_2(\theta_q) \rho^2\right) e^{\rho\cos\theta_q} \nonumber \\
&+ C_1(\theta_q) e^{\rho\cos\theta_q} +\frac{1}{2\pi},
\end{align}
where
\begin{align}
&B_1 = -A_1\sec^2\theta_q - \tan\theta_q \frac{d A_1}{d\theta_q},\\
&B_2 = \frac{1}{2}A_1\sin\theta_q\tan\theta_q
\end{align}
and the far-field condition is already enforced. Making use of the no-flux condition at $\rho=0$, we obtain an ordinary differential equation (ODE) for $A_1$,
\begin{align}
\frac{\cos\theta_q}{2\pi} +A_1 \sec^2\theta_q +\tan\theta_q \frac{d A_1}{d\theta_q}=0.
\end{align}
Requiring regularity of $A_1$ at $\theta_q = \pi$, we can integrate the above equation to obtain $A_1 = -\cos\theta_q/(2\pi)$. 

In Fig. \ref{fig:g0m1}(a)  we plot the leading-order probability distribution $g_0^{(-1)}$  as a function of $\rho$ and $\theta_q$. In Fig. \ref{fig:g0m1}(b) , we show the number density $n_0^{(-1)} = \int g_0^{(-1)}d\theta_q$ and the radial polar order $m_{\rho,0}^{(-1)}= \int g_0^{(-1)}\cos\theta_qd\theta_q$ as a function of $\rho$. In the boundary layer, ABPs are pointing into the probe, because otherwise they would swim away. As a result, we observe an accumulation at contact ($\rho=0$) and a negative radial polar order. The following asymptotic behaviors can be obtained near contact:
\begin{align}
n_0^{(-1)}  &= \frac{1}{\pi} -\frac{\rho}{4} +O(\rho^2) \quad \text{as}\quad \rho \to 0,\\ 
m_{\rho,0}^{(-1)} &= -\frac{1}{4}+ \frac{2\rho}{3\pi}+O(\rho^2)\quad \text{as}\quad \rho \to 0.
\end{align}
It is worth noting that the boundary layer structure is identical in 2D and 3D. One can show that in 3D the leading-order probability density is given by $g_0^{-1} = \bq\cdot\be_r \exp(\bq\cdot\be_r \rho)/(-8\pi)$, which differs from \eqref{eq:CV-g0m1} only by a numerical factor due to the dimensionality.

To determine the function $C_1$ in Eq. \eqref{eq:CV-slow-probe-g00}, we again need the solution at the next order. So far in this section we have considered the asymptotic behavior of the probability distribution of ABPs outside of a fixed probe in the large $\Pe_s$ limit. This problem  has been solved by \citet{yan_brady_2015} using a $\bQ=\bm{0}$ [defined by $\bQ=\int g(\bq\bq-\bI/d)d\bq$] closure and BD simulations. We note that the $\bQ$ closure gives the correct scaling for the number density, i.e., the number density at contact scales as $Pe_s$ for large $Pe_s$, but does not give quantitatively correct results over the full range of $Pe_s$.

We now consider the microstructure disturbance due to the weak probe motion and the zero-velocity microviscosity in the high $Pe_s$ limit. The governing equations for $f_1$ and $f_2$ in the boundary layer can be obtained similarly to the approach described above for $g_0$. At leading order, the disturbance fields are finite, i.e., $O(\epsilon^0)$, and are expanded as (recall that $\epsilon=1/Pe_s$)
\begin{align}
&f_1(\rho, \theta_q) = f_1^{(0)} + \epsilon f_1^{(1)} +O(\epsilon^2),\\
&f_2(\rho, \theta_q) = f_2^{(0)} + \epsilon f_2^{(1)} +O(\epsilon^2).
\end{align}
The governing equation for $f_1^{(0)}$ is 
\begin{align}
&\cos\theta_q \frac{\partial f_1^{(0)}}{\partial \rho} - \frac{\partial^2 f_1^{(0)}}{\partial \rho^2} = \frac{\partial g_0^{(-1)}}{\partial \rho},\\
&\cos\theta_q f_1^{(0)} - \frac{\partial f_1^{(0)}}{\partial \rho} = g_0^{(-1)} \quad\text{at}\quad \rho=0,\\
& f_1^{(0)} \to 0\quad\text{as}\quad \rho \to \infty.
\end{align}
The solution is given by 
\begin{align}
\label{eq:CV-f1-leading-sol}
f_1^{(0)} = \frac{1}{2\pi}\rho\cos\theta_q  e^{\rho\cos\theta_q} +A_2(\theta_q) e^{\rho\cos\theta_q},
\end{align}
which is only valid for $\cos\theta_q <0$. To determine $A_2$, we need the solutions to $f_2^{(0)}$ and  $f_1^{(1)}$. Imposing regularity of $f_1^{(0)}$ at $\theta_q=\pi$, one can show that $A_2 = 1/(2\pi)$. Finally, using Eq. \eqref{eq:CV-eta-0}, we obtain the zero-velocity microviscosity in the fast-swimming limit,
\begin{equation}
    \eta_0^\mathrm{micro} \to \frac{1}{2}\quad\text{as}\quad Pe_s \to \infty.
\end{equation}

\subsection{Zero-velocity microviscosity}
To obtain the microstructure in the zero $Pe$ limit for arbitrary values of $Pe_s$, we solve Eqs. \eqref{eq:CV-slow-probe-g0}--\eqref{eq:Pe-1-bc-f2} numerically using a Fourier--Laguerre spectral method (see Sec. \ref{sec:cv-numerics}). For large $Pe_s$, the discretization of the equations needs to conform with the boundary layer structure as discussed in Sec. \ref{sec:fast-swimming} in order to yield accurate numerical results. To this end, for $Pe_s  > 10$, instead of discretizing $r$, the boundary layer coordinate $\rho$ is discretized and used in the numerical solution.

\begin{figure}
  \centering
  \includegraphics[width=\columnwidth]{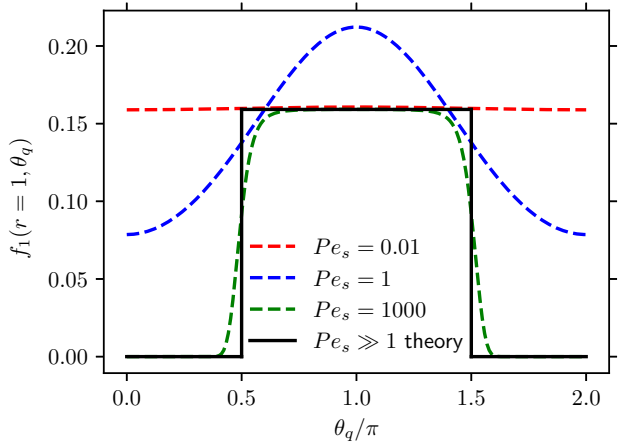}
  \caption{\label{fig:f1-contact} The function $f_1(r=1, \theta_q)$ as a function of $\theta_q$ for several values of $Pe_s=U_2R_c/D_2$ and $\gamma=R_c/\delta=1$. The large $Pe_s$ asymptotic solution given by Eq. \eqref{eq:CV-f1-leading-sol} is plotted as a solid line. The zero-velocity microviscosity is the area under the curve as can be seen from Eq. \eqref{eq:CV-eta-0}. }
\end{figure}

As shown in Eq. \eqref{eq:CV-eta-0}, the contact distribution of $f_1$ determines the zero-velocity microviscosity. More precisely, $\eta_0^\text{micro}$  is the area under the curve  $f(r=1, \theta_q)$ from $\theta_q=0$ to $\theta_q=2\pi$. For a passive suspension, $Pe_s \equiv 0$, one can readily show that $f_1(r=1,\theta_q) \equiv 1/(2\pi)$, in which case the area under the curve is unity and hence $\eta_0^\text{micro} = 1$.  For large $Pe_s$, the contact distribution of $f_1$ obtained from Eq. \eqref{eq:CV-f1-leading-sol} is $f_1(r=1, \theta_q) = 1/(2\pi)$ for $\bq\cdot\be_r <0$ and zero otherwise. In other words, the contact value of $f_1$ for large $Pe_s$ is the same as the limit of $Pe_s \to 0$ but only in half of the domain of $\theta_q$. Therefore, the zero-velocity microviscosity approaches $1/2$ as $Pe_s \to \infty$. In Fig. \ref{fig:f1-contact} we plot the contact distribution of $f_1$ as a function of $\theta_q$ for several values of $Pe_s$ obtained from the numerical solutions. The leading-order asymptotic solution in the large $Pe_s$ limit is plotted as a solid line.

\begin{figure}
  \centering
  \includegraphics[width=\columnwidth]{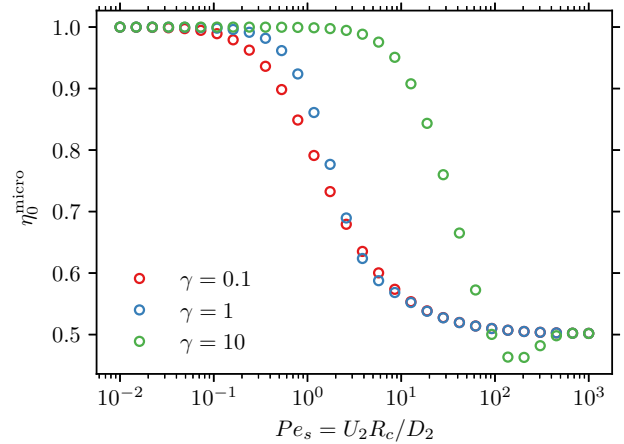}
  \caption{\label{fig:zero-Pe-eta}  The zero-velocity microviscosity $\eta_0^\text{micro}$ as a function of $Pe_s=U_2R_c/D_2$ for several values of $\gamma=R_c/\delta$.}
\end{figure}

In Fig. \ref{fig:zero-Pe-eta} we present the zero-velocity microviscosity as a function of $Pe_s$ for several values of $\gamma$. As alluded to earlier, the zero-velocity microviscosity exhibits a swim-thinning behavior. That is, $\eta_0^\text{micro}$ in general decreases with increasing swim speed, or $Pe_s$. The onset of swim-thinning occurs at $O(Pe_s^2)$ for small $Pe_s$ (see appendix \ref{sec:slow-swimming}). An outlier in this general behavior appears when $\gamma^2$ is comparable to $Pe_s$ and both are large, $Pe_s \sim \gamma^2 \gg 1$. This can be seen from the results in Fig. \ref{fig:zero-Pe-eta} for $\gamma=10$, in which case $\eta_0^\text{micro}$ decreases below $1/2$ before increasing and asymptoting to the large $Pe_s$ value of $1/2$. An asymptotic analysis in the limit $Pe_s \gg 1$ while $Pe_s/\gamma^2 =O(1)$ shows that the boundary layer thickness remains the same but an additional forcing term due to rotary diffusion appears in Eq. \eqref{eq:CV-slow-probe-g00-eq}. The addition of this new term renders the boundary layer equations analytically intractable.

\section{Numerical solution of the Smoluchowski equation}
\label{sec:cv-numerics}
To obtain the suspension microstructure over the full range of $Pe$, a numerical solution of the full Smoluchowski Eq. \eqref{eq:CV-smol-g-2D-explicit-eq} together with its boundary conditions \eqref{eq:no-flux-g-2d-explicit} and \eqref{eq:g-2d-explicit-far} is required. In this section, we develop a Fourier--Laguerre spectral method in which the physical space angular position $\theta$ and the orientation angle $\theta_q$ are resolved analytically using a truncated double Fourier series expansion. To this end, we first approximate the microstructure at steady state as a truncated double Fourier series as 
\begin{align}
\label{eq:CV-g-double-fourier}
    g(r, \theta, \theta_q) \approx \sum_{m=-M}^{M}\sum_{n=-N}^NC_{m,n}(r)e^{im\theta}e^{i n \theta_q},
\end{align}
where $i^2=-1$ is the imaginary unit and $C_{m,n}(r)$ is the Fourier mode indexed by $m$ and $n$.  Inserting Eq. \eqref{eq:CV-g-double-fourier} into \eqref{eq:CV-smol-g-2D-explicit-eq}, at steady state we obtain a system of $(2N+1)(2M+1)$ coupled ODEs for the radially varying Fourier modes,
\begin{align}
\label{eq:Fourier-expanded-ODEs}
    &\frac{Pe_s}{2}\frac{d }{dr}\left( C_{m,n+1} +C_{m,n-1}\right) \nonumber\\
    &+ \frac{Pe_s}{2r}\left[ (n+1-m)C_{m,n+1} + (m-(n-1)C_{m,n-1}\right]\nonumber \\
    &-\frac{Pe}{2}\frac{d}{dr}\left(C_{m+1,n} +C_{m-1,n} \right) \nonumber \\
    &+\frac{Pe}{2r}\left[(n-(m+1)C_{m+1,n} +(m-1-n)C_{m-1,n} \right]\nonumber \\
    & -\frac{1}{r}\frac{d}{dr}r\frac{dC_{m,n}}{dr} + \frac{m^2+n^2-2mn}{r^2}C_{m,n} +\gamma^2 n^2 C_{m,n}=0.
\end{align}
Here, any Fourier mode $C_{m,n}$ that exceeds the range $-M \leq m \leq M, -N \leq n \leq N$ is simply discarded. Similarly, the no-flux condition \eqref{eq:no-flux-g-2d-explicit} becomes 
\begin{align}
\label{eq:CV-no-flux-Fourier}
    &\frac{Pe_s}{2}\left( C_{m,n+1}+ C_{m,n-1}\right) \nonumber\\
    &- \frac{Pe}{2}\left( C_{m+1,n}+C_{m-1,n}\right) - \frac{d C_{m,n}}{dr}=0\quad\text{at}\quad r=1,
\end{align}
and the far-filed condition \eqref{eq:g-2d-explicit-far} is 
\begin{equation}
  C_{m,n} \to
    \begin{cases}
      \frac{1}{2\pi} & m=n=0\\
      0 & \text{otherwise}
    \end{cases}       \quad\text{as}\quad r\to \infty.
\end{equation}

\begin{figure*}
  \centering
  \includegraphics[width=5in]{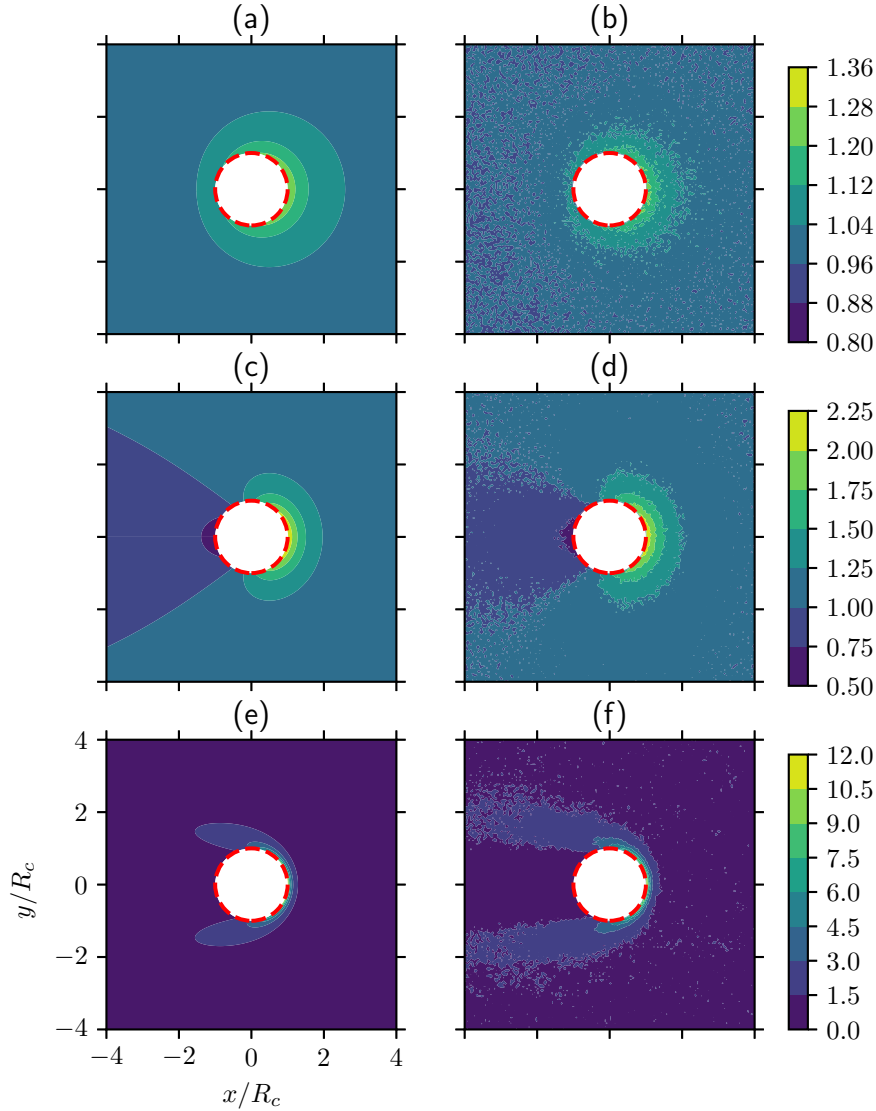}
  \caption{\label{fig:cv-density-contour} Contour plots of the number density distribution around the probe for different values of $Pe=U_1R_c/D_2$ with $Pe_s=U_2R_c/D_2=1, \gamma=R_c/\delta=1$ obtained from the numerical solution of the Smoluchowski equation [(a), (c), and (e)] and BD [(b), (d), and (f)]. For the top panels (a) and (b), $Pe=0.1$; for (c) and (d), $Pe=1$, and for (e) and (f), $Pe=10$. All panels have identical $x$ and $y$ limits and are thus only shown in (e). Panels in each row have the same color bar and are shown on the right. The red disk with a white fill represents the circle of contact with radius $R_c$. }
\end{figure*}

\begin{figure}
    \centering
    \includegraphics[width=3in]{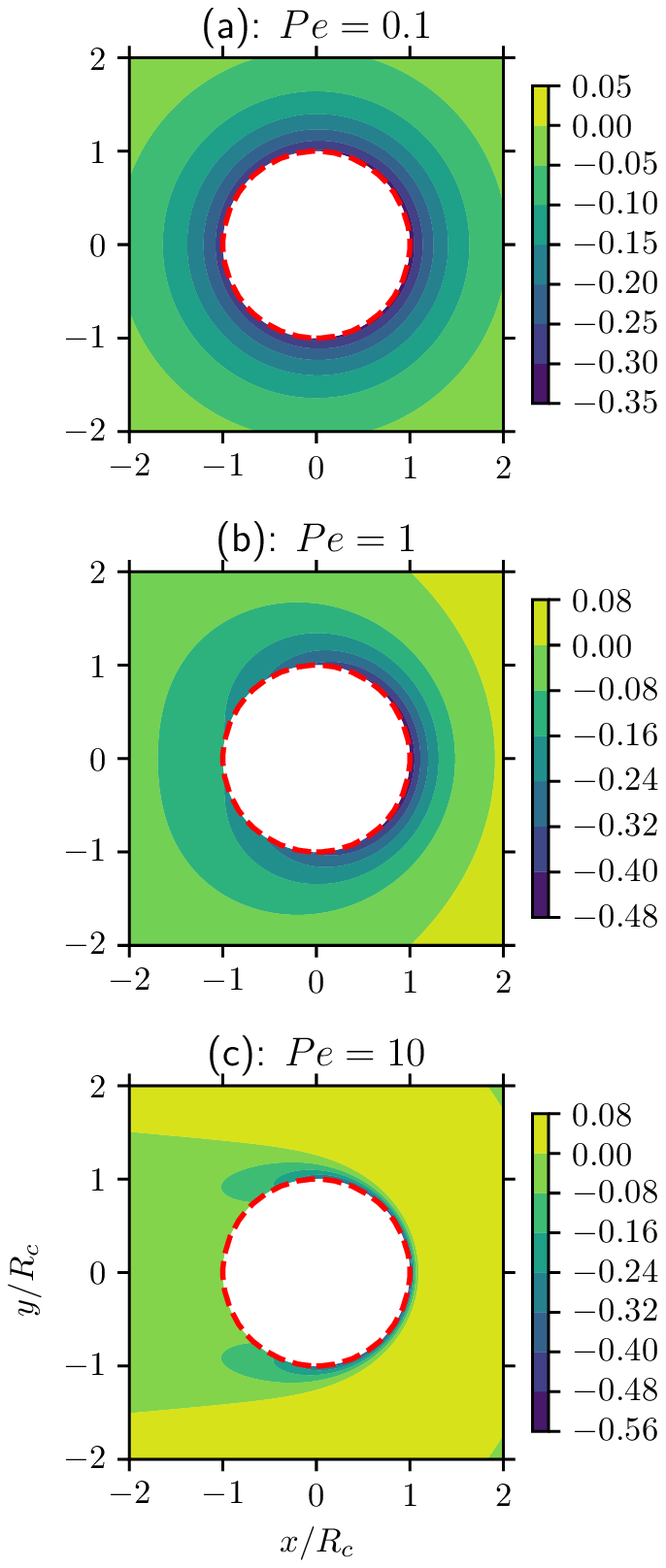}
    \caption{\label{fig:cv-polar-contour} Contour plots of the radial polar order distribution ($m_r = \int g \bq\cdot \be_r d\bq $) around the probe for different values of $Pe=U_1R_c/D_2$ with $Pe_s=U_2R_c/D_2=1, \gamma=R_c/\delta=1$ obtained from the numerical solution of the Smoluchowski equation. All panels have identical $x$ and $y$ limits and are thus only shown in (c).  The red disk with a white fill represents the circle of contact with radius $R_c$. The contact values of the radial polar order, $m_r(r=1, \theta)$, are plotted in Fig. \ref{fig:cv-contact-n-mr}(b).}
  \end{figure}

We solve the system of ODEs in \eqref{eq:Fourier-expanded-ODEs} using spectral collocation of the Laguerre functions at the Laguerre--Gauss--Radau quadrature nodes \citep{shen2011spectral}. The Laguerre function of order $n$ is defined by $\hat{L}_n(x) = e^{-x/2}L_n(x)$, where $L_n(x)$ is the Laguerre polynomial satisfying the recurrence relation $(n+1)L_{n+1} = (2n+1-x)L_n- n L_{n-1}$  and $L_0(x)=1, L_1(x) = 1-x$. The orthogonality condition of the Laguerre functions is given by $\int_0^{+\infty} \hat{L}_j(x) \hat{L}_k(x) d x = \delta_{jk}$ where $\delta_{jk}$ is the Kronecker delta. It is clear that all Laguerre functions vanish at infinity. To accommodate this natural boundary condition, we define the shifted Fourier modes $\tilde{C}_{m,n}$ such that $\tilde{C}_{0,0} = C_{0,0}-1/(2\pi)$ and $\tilde{C}_{m,n}  =C_{m,n}$ otherwise. It is straightforward to rewrite the ODEs in \eqref{eq:Fourier-expanded-ODEs} and the no-flux condition \eqref{eq:CV-no-flux-Fourier} in terms of $\tilde{C}_{m,n}$.  For $Pe, Pe_s \lesssim 10$, the ODEs in terms of $\tilde{C}_{m,n}$ are solved after shifting the radial coordinate $r \to r-1$ so that it falls into the natural domain $[0, \infty)$ of the Laguerre functions. For $Pe_s \gtrsim 10$, there exists an accumulation boundary layer near the wall as considered in Sec. \ref{sec:slow-probe}; in this case, a stretched coordinate $\rho = (r-1)Pe_s$ is used and the ODEs are written in terms of $\rho$ before applying the spectral collocation. For a fast-moving probe, $Pe \gg 1$, there exists a boundary layer of thickness $O(1/Pe)$ in the front sector of the probe with density in the boundary layer growing like $Pe$ as $Pe \to \infty$, just like the case of a probe moving in a passive Brownian suspension \citep{SquiresBrady2005}. For $Pe \gtrsim 10$, we use the stretched coordinate $\rho = (r-1)Pe$ for the spectral collocation.

To obtain the zero-velocity microviscosity discussed in Sec. \ref{sec:slow-probe}, Eqs. \eqref{eq:CV-slow-probe-g0}--\eqref{eq:Pe-1-bc-f2} are solved numerically using the above-mentioned approach. The results obtained by solving \eqref{eq:CV-slow-probe-g0}--\eqref{eq:Pe-1-bc-f2} agree with the numerical solution of the full Smoluchowski equation with a small $Pe$ number.

Because the resulting discretized linear system has a very large dimension and the spectral differentiation matrix is dense, the matrix system is not formed explicitly. We solve the linear system iteratively using a matrix-free generalized minimal residual method.

In Fig. \ref{fig:cv-density-contour} we plot the number density distribution (recall  that $n = \int g d\bq$) in a region around the probe for several values of $Pe$ with $Pe_s=1$ and $\gamma=1$. Contours on the left [(a), (c), (e)] are obtained from the numerical solutions of the Smoluchowski equation and are compared to the results obtained from BD (see Sec. \ref{sec:CV-BD} for details on BD) on the right [(b), (d), (f)]. The $Pe$ numbers from the top to the bottom are, respectively, $0.1$, $1$, and $10$. A square grid is used to sample the number density distribution and is averaged over several hundred frames at long times. Despite the noise, the density distribution sampled from BD agrees well with that obtained from solving the Smoluchowski equation. 

It is important to note that the suspension microviscosity defined by Eqs. \eqref{eq:eta-micro-3d} and \eqref{eq:eta-micro-2d} only depends on the number density of bath particles at contact. In contrast to passive bath particles, activity (i.e., self-propulsion) only provides an additional mechanism for the transport of bath particles---they self-propel. By integrating out the orientational degrees of freedom from the Smoluchowski Eq. \eqref{eq:CV-smol-g}, one obtains the governing equation for the number density as $Pe \bUh_1\cdot\nabla_r n + \nabla_r^2 n  = Pe_s \nabla_r\cdot\polarorder$, where $\polarorder = \int \bq g d\bq$ is the polar order. One can see that the polar order $\polarorder$ serves as a forcing term in the governing equation for the number density $n$. In Fig. \ref{fig:cv-polar-contour} we plot the radial polar order distribution ($m_r = \be_r \cdot \polarorder$) corresponding to the density profiles shown in Fig. \ref{fig:cv-density-contour}.

\begin{figure*}
  \centering
  \includegraphics[width=6in]{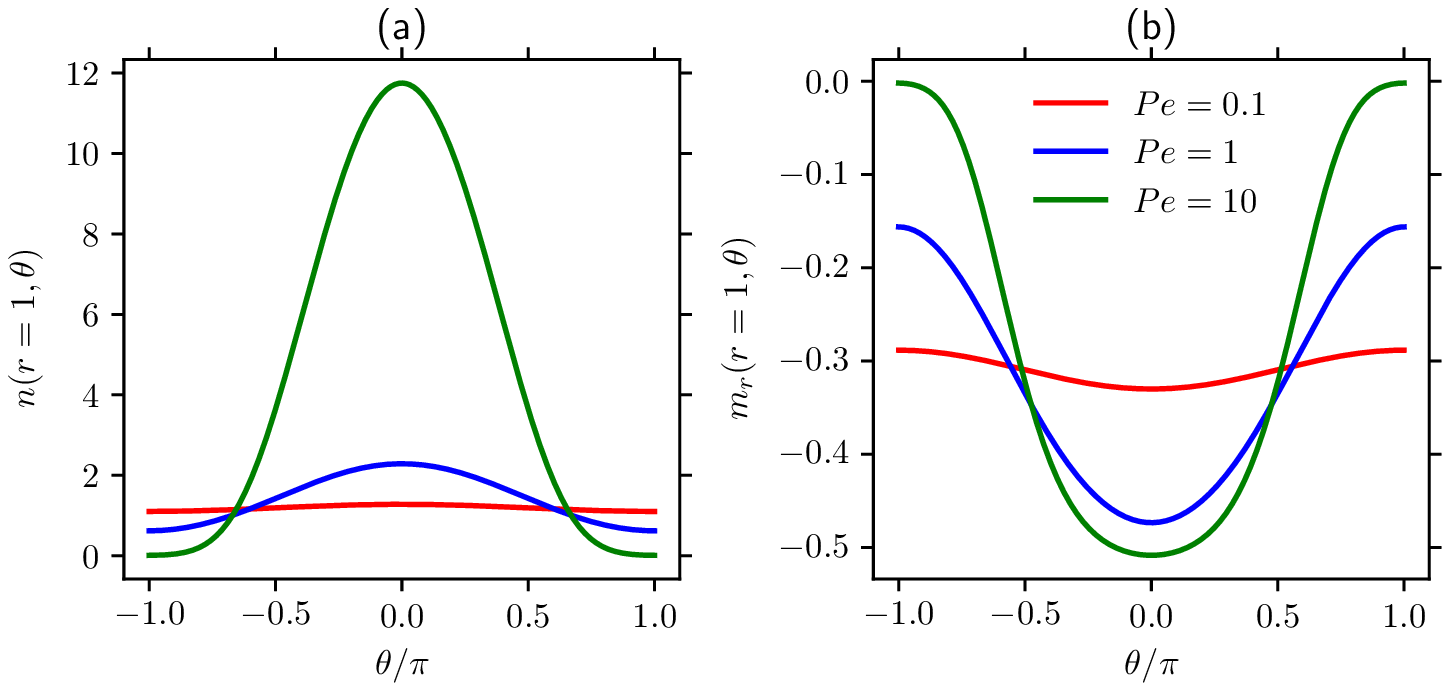}
  \caption{\label{fig:cv-contact-n-mr} Contact values of (a) the number density and (b) the radial polar order as a function of the angular position $\theta$ for several values of $Pe=U_1R_c/D_2$. Panels (a) and (b) share the same legends and are only shown in (b). For all lines plotted, $Pe_s = U_2R_c/D_2=1$ and $\gamma = R_c/\delta = 1$. The front of the probe is at $\theta=0$ and the back is at $\theta=\pm \pi$. }
\end{figure*}

The number density and radial polar order distributions at contact corresponding to the contour plots shown in Figs. \ref{fig:cv-density-contour} and \ref{fig:cv-polar-contour} are shown in Fig. \ref{fig:cv-contact-n-mr}.

When the speed of the probe is zero, i.e., $Pe=0$, the microstructure (hence the density) is isotropic, which is simply the distribution of ABPs outside of a fixed disk \citep{yan_brady_2015}. Because the suspension is active, the density is not uniform in space but exhibits a boundary accumulation at contact. In order for ABPs to accumulate at the boundary, they must exhibit a net polar order pointing into the boundary ($m_r <0$) because otherwise they swim away. In the absence of activity ($Pe_s=0$), the number density is uniform. When the probe is set into motion in an active suspension, the microstructure is perturbed from its isotropic steady state. (For an active suspension, this steady state is not in  thermodynamic equilibrium.) For small $Pe$ such as that shown in Fig. \ref{fig:cv-density-contour}(a), the microstructure is only  slightly perturbed from the isotropic state and has been characterized in section \ref{sec:slow-probe}. As $Pe$ increases, a prominently nonuniform density distribution develops at contact with an accumulation at the front and a depletion in the back of the probe as can be seen in Fig. \ref{fig:cv-contact-n-mr}(a). Because the density becomes depleted in the back, the polar order also decreases in the back (in absolute value) and increases in the front of the probe as shown in Fig. \ref{fig:cv-contact-n-mr}(b).

\begin{figure}
  \centering
  \includegraphics[width=\columnwidth]{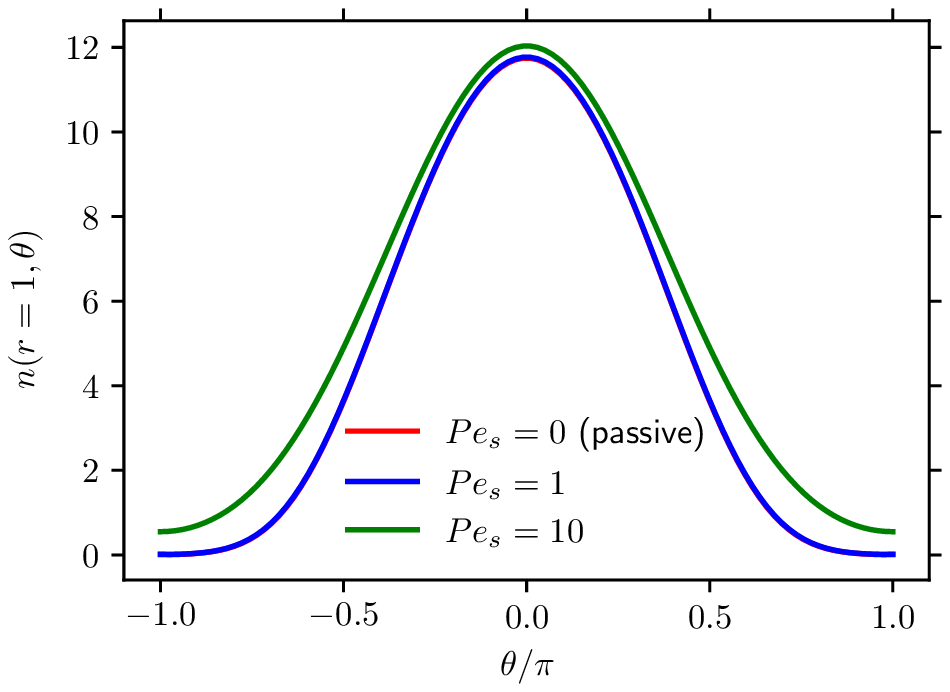}
  \caption{\label{fig:cv-contact-n-vs-Pes} Number density at contact for several values of $Pe_s = U_2R_c/D_2$. For all results,  $\gamma=Rc/\delta=1$ and $Pe=U_1R_c/D_2=10$. The blue and red lines agree. }
\end{figure}

For passive Brownian suspensions, the buildup of particles in the front of the probe is solely due to the advection of the probe. When the suspension is active, this advective effect is still present. In Fig. \ref{fig:cv-contact-n-vs-Pes} we plot the contact density at $Pe=10$ for a passive suspension and for active suspensions with $Pe_s=1$ and $Pe_s=10$. For $Pe_s=1$, the swim speed is small compared to that of the probe and the contact density is almost the same as that of the passive suspension. For $Pe_s=10$, the swim speed is comparable to the probe speed and the density is elevated from that of the passive. This elevation in density represents the additional wall accumulation resulting from activity. In particular, we note that the contact density on all sides of the probe is increased.

\section{BD simulation}
\label{sec:CV-BD}
From a particle-level perspective, the evolution of the configuration of ABPs can be described by the overdamped Langevin equations---a balance of forces and torques. In the absence of HIs as we consider here, BD can be used to simulate the dynamics of ABPs at the particle level. BD has been used to study the bulk rheology \citep{Foss2000BD} and microrheology \citep{Carpen2005Micro} of passive colloidal suspensions. Our approach is similar to those considered by \citet{Foss2000BD} and \citet{Carpen2005Micro} except that the orientational dynamics of each particle also needs to be tracked due to the self-propulsion of ABPs.

 For each ABP, the force and torque balance in the comoving frame is given by 
\begin{align}
\label{eq:CV-langevin-translation}
 \bm{0} &= -\zeta_2\left( \frac{d \bx }{dt} + \bU_1\right) + \bF^B +\bF^S + \bF^{HS}, \\
 \label{eq:CV-langevin-rotation}
 \bm{0} &= -\zeta_R \frac{d\bq}{dt} + \bL^B\times \bq.
\end{align}
Here, $\bF^S = \zeta_2U_s\bq$ is the swim force \citep{takatoriprl14} giving rise to self-propulsion, $\bF^B$ ($\bL^B$) is the Brownian force (torque), $\bF^{HS}$ is the hard-sphere force due to the steric interaction between the probe and the ABP, and $\zeta_R$ is the rotational hydrodynamic drag coefficient. Because both the probe and the ABP are spheres, their hard-sphere interaction does not induce a torque. However, if either the probe or the ABP (or both) is nonspherical, their hard-particle interaction can induce a torque. The hard-sphere force is present only when the probe and the ABP are in contact and is the mechanism of the enhanced viscosity in passive colloidal suspensions compared to the motion of a probe in the solvent alone. The Brownian force and torque satisfy the white-noise statistics,
\begin{align}
    \left< \bF^B\right> &= \bm{0}, \quad \left< \bF^B(0)\bF^B(t)\right> = 2 D_2 \zeta_2^2\delta(t) \bI,\\
    \left< \bL^B\right> &= \bm{0}, \quad \left< \bL^B(0)\bL^B(t)\right> = 2 D_R \zeta_R^2\delta(t) \bI,
\end{align}
where $\delta(t)$ is the delta function (which has the units of the inverse of time) and the angle brackets denote the ensemble average over Brownian fluctuations. We emphasize that the rotational diffusivity represents biological noises and can be varied independently from the translational diffusivity $D_2$. In the comoving frame, the probe is fixed in space and appears as an obstacle for the dynamics of ABPs outside of it.

In 2D, using  the parametrization $\bq(t) = \cos\theta^\prime(t) \be_x +\sin\theta^\prime(t) \be_y$, it is straightforward to see that $d\bq/dt = \be_z\times \bq d\theta^\prime/dt$ where $\be_z = \be_x\times \be_y$. As a result, Eq. \eqref{eq:CV-langevin-rotation} can be written as 
\begin{equation}
\label{eq:CV-langevin-angular-reduced}
    \frac{d\theta^\prime }{dt} = \Omega^B,
\end{equation}
where the Brownian angular velocity satisfies $\left<\Omega^B(t) \right> =0$ and $\left< \Omega^B(0)\Omega^B(t)\right> = 2D_R \delta(t)$.

Using the Euler--Maruyama scheme, the linear and angular Eqs. \eqref{eq:CV-langevin-translation} and \eqref{eq:CV-langevin-angular-reduced} can be  discretized given the time step $\Delta t $; their discrete forms at time $t=t_k = k\Delta t (k=0,1,...)$ are given by 
\begin{align}
\label{eq:CV-langevin-translation-discret}
    \bx_{k+1}&= \bx_k + \left( -\bU_1 + U_s \bq_k  \right)\Delta t + \sqrt{2D_2\Delta t}\bm{\xi}_x +\Delta \bx^{HS},\\ 
    \label{eq:CV-langevin-rotation-discret}
    \theta^\prime_{k+1} &=\theta^\prime_k + \sqrt{2D_R\Delta t}\xi_{\theta^\prime}, 
\end{align}
where $\bm{\xi}_x$ is a two-vector of pseudo-random numbers with each entry having zero mean and unit variance. Similarly, $\xi_{\theta^\prime}$ is a scalar having zero mean and unit variance.

At each time step, the position of the ABP is updated first by adding the displacements due to the relative velocity $-\bU_1$, the swimming, and the Brownian contributions and second by resolving collision with the fixed probe. We use the potential-free algorithm \citep{Heyes1993,Foss2000BD} in which the overlap between the probe-ABP pair is corrected by moving the ABP along the line of centers back to contact. Because the probe has prescribed kinematics (i.e., fixed velocity), only the ABP is moved if an overlap is detected. 

At this point, a contrast between the CF and CV modes of microrheology is in order. In the CF mode of microrheology, either the external force is zero (tracer dispersion) or finite, and in the collision resolution step, both the probe and the ABP have to be displaced in opposite directions such that Newton's third law is satisfied. In the CV mode of microrheology, because the probe is never displaced due to collision, one can have many bath particles interacting with a single probe in one simulation; these bath particles are ``transparent'' to each other in the sense that they can pass through each other and only interact with the probe.  For the CF mode, however, the collision resolution between the probe and a bath particle might introduce a new overlap between the probe and a different bath particle due to the displacement of the probe. Therefore, if one wishes to simulate the pair-interaction between the probe and one bath particle only, one can run many independent simulations each consisting of the probe and a bath particle or simulate a system of many bath particles with low volume fraction. The results obtained in the second method is a good approximation to the pair behavior only when the system is sufficiently dilute.

In the simulation setup, the system consists of the fixed probe and $N$ ABPs in a rectangular domain of lengths $L_x$ and $L_y$, where the $x$ axis is aligned with $\bU_1$, i.e., $\bUh_1 = \be_x$. The size of the simulation domain needs to be sufficiently large such that its boundary is a good approximation of the far-field [see Eq. \eqref{eq:CV-far-field-g}]. In particular, the domain needs to be much larger than the run length of the ABPs. At each time step, we evolve the positions and orientations of all ABPs according to Eqs. \eqref{eq:CV-langevin-translation-discret} and \eqref{eq:CV-langevin-rotation-discret} and the hard-sphere displacement of each particle is recorded when necessary. Simulations are performed using an in-house CUDA-accelerated code that runs on NVIDIA GPUs, which enables us to run a typical simulation with $O(10^5)$ ABPs. The measured area fraction of the ABPs is $\phi = N \pi b^2 / (L_xL_y - \pi a^2)$. Because the ABPs are transparent to each other, the measured area fraction has no physical interpretation and only serves to improve the measured statistics.

The force balance of the probe is $-\zeta_1 \bU_1 +\bF^\text{ext} - \bF^{HS}=\bm{0}$, where the hard-sphere force the ABP exerts on the probe is $-\bF^{HS}$ according to Newton's third law. To maintain a CV of the probe, the external force fluctuates and we are concerned with its average over the fluctuations. Noting that $\bF^{HS} = \zeta_2 \Delta\bx^{HS}/\Delta t$ \citep{Foss2000BD} and Eq. \eqref{eq:CV-viscosity-increment}, we obtain 
\begin{equation}
    \frac{\Delta \eta}{\eta} = \frac{\zeta_2 \Delta x^{HS}}{\zeta_1 U_1 \Delta t},
\end{equation}
where $\Delta x^{HS}$ is the accumulated hard-sphere displacement of all $N$ ABPs at each time step and then averaged over many frames at sufficiently long times so that a steady state is reached. It is then straightforward to calculate $\eta^\text{micro}$ using the first part of Eq. \eqref{eq:eta-micro-2d}. 

\begin{figure*}
\centering
\includegraphics[width=5in]{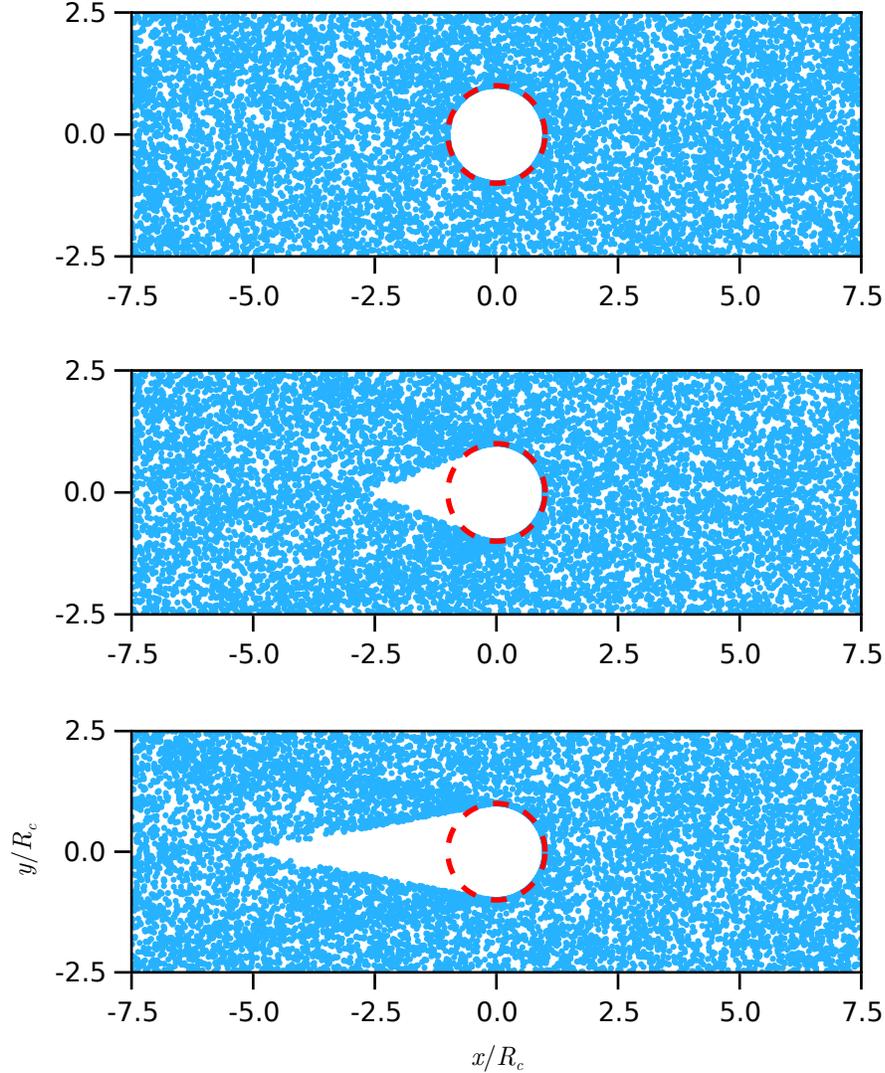}
\caption{\label{fig:bd-snapshots} Windowed snapshots of the BD simulation  showing the circle of contact (red dashed line), the positions of ABPs (blue dots), and the wake structure behind the probe. In all panels, $\ell/R_c = 1$ and $D_2 =0$. The speed of the probe increases from top to bottom: $U_1\tau_R/R_c = \{0.5, 2.5, 5\}$. The blue dots denote the center positions of the ABPs and their size in the figure does not represent the size of the ABPs in the simulation. The simulation domain is larger than the window shown and $10000$ ABPs are plotted in each panel. }
\end{figure*}

In Fig. \ref{fig:bd-snapshots} we show three snapshots of the BD simulation for varying probe speeds. The speed of the probe increases from the top panel to the bottom.  The snapshot is windowed around the probe in order to highlight the near-field microstructure. The red dashed circle denotes the circle of contact that is concentric with the probe but with radius $R_c$ (see Fig. \ref{fig:cvmicro-schematic}). The blue dots are the positions of the centers of the ABPs and their size does not reflect the size of ABPs in the simulation. A prominent feature of the near-field microstructure is the presence of a trailing wake behind the probe that is devoid of bath particles. To highlight the wake structure, in Fig. \ref{fig:bd-snapshots} the translational diffusion is turned off, $D_2 =0$. In the absence of translational diffusion, the only source of noise in the dynamics of ABPs comes from the Brownian reorientation. Recall that in the simulation the probe is fixed in place while the ABPs experience a constant advection of speed  $U_1$ to the left. To understand the development of the trailing wake, consider an ABP that is behind the probe (to the left). In order for this ABP to reach the probe from behind, the best orientation it should take is $\bq = \be_x$ (pointing to the right), in which case the net speed to the right is $U_2-U_1$. When $U_2 > U_1$, i.e., the speed of the ABPs is larger than that of the probe, ABPs with orientations near $\be_x$ can reach the probe. On the other hand, when $U_2 \leq U_1$, the speed of the ABPs is smaller than that of the probe and all ABPs will be advected further to the left. This simple physical argument suggests that the onset of a trailing wake happens when $U_2 \approx U_1$. Indeed, when the speed of the probe is smaller than that of the ABPs as shown in the top panel of Fig. \ref{fig:bd-snapshots} there is no wake. When the speed of the probe surpasses that of the ABPs, e.g. in the middle and bottom panels of Fig. \ref{fig:bd-snapshots}, the trailing wake appears and becomes more extended to the left as the speed of the probe increases.

To characterize the geometry of the triangular wake, we define the wake half angle $\beta$, which is the angle between the top (or equivalently the bottom) boundary of the wake and the horizontal axis. Consider an active particle  in contact with the top of the probe; in order for this particle to swim into the back of the probe, it should have an orientation toward the bottom. The maximum vertical displacement is achieved with $\bq = -\be_y$, in which case the ABP assumes a trajectory that has a slope given by $\tan \beta = U_2/U_1$. This simple argument is able to predict the wake half angle quantitatively. The wake half angle can be easily read off from the middle and bottom panels of Fig. \ref{fig:bd-snapshots}. Taking the bottom panel as an example, from the plot, we see that $\tan\beta \approx 1/5$, which is exactly the speed ratio $U_2/U_1=1/5$. Similarly, one can verify the prediction in the middle panel.

In the presence of translational diffusion, the wake boundary becomes less sharp because of the diffusive flux $-D_2 \nabla n$ down the number density gradient, into the wake. If the swim speed is small compared to the probe speed, the wake structure with finite $D_2$ approaches that obtained for a passive Brownian suspension, which has been studied \citep{SquiresBrady2005, Carpen2005Micro}.

In Fig. \ref{fig:bd-snapshots}, periodic boundary conditions in both directions of the simulation domain are used. This is a good approximation of the far-field condition provided that the simulation domain is sufficiently long such that all prior interactions of an ABP with the probe have relaxed once the ABP reaches the boundary of the domain. When the probe speed is much larger than that of the ABPs, $U_1 \gg U_2$, the trailing wake becomes rather extended in the horizontal direction. To make use of the periodic boundary condition, the simulation domain has to be enlarged accordingly, which makes the nominal volume fraction very small and the number of particles colliding with the probe diminishing. For a fixed number of ABPs, as the probe speed increases, the statistics for the hard-sphere force becomes less reliable. Note that for $U_1 \gg U_2$, once an ABP moves past the probe to the left, the chance of it turning back to the right without exiting the left boundary and entering from the right is vanishingly small. As a result, we introduce a new boundary condition in the horizontal direction for $U_1\gg U_2$ in which the trailing wake is cut off. Once an ABP moves past the probe to the left, it is removed from the simulation and added back from the right boundary of the domain with the bulk distribution, i.e., random orientation and $y$ position. Similarly, if the ABP leaves the boundary from the right, instead of appearing from the left, it will be put back on the right boundary with the bulk distribution. This new boundary condition allows us to reduce the domain size significantly for $U_1\gg U_2$ but still obtain the correct microviscosity measurement.

\section{Microviscosity}
\label{sec:cv-eta-micro}
\begin{figure*}
  \centering
  \includegraphics[width=6in]{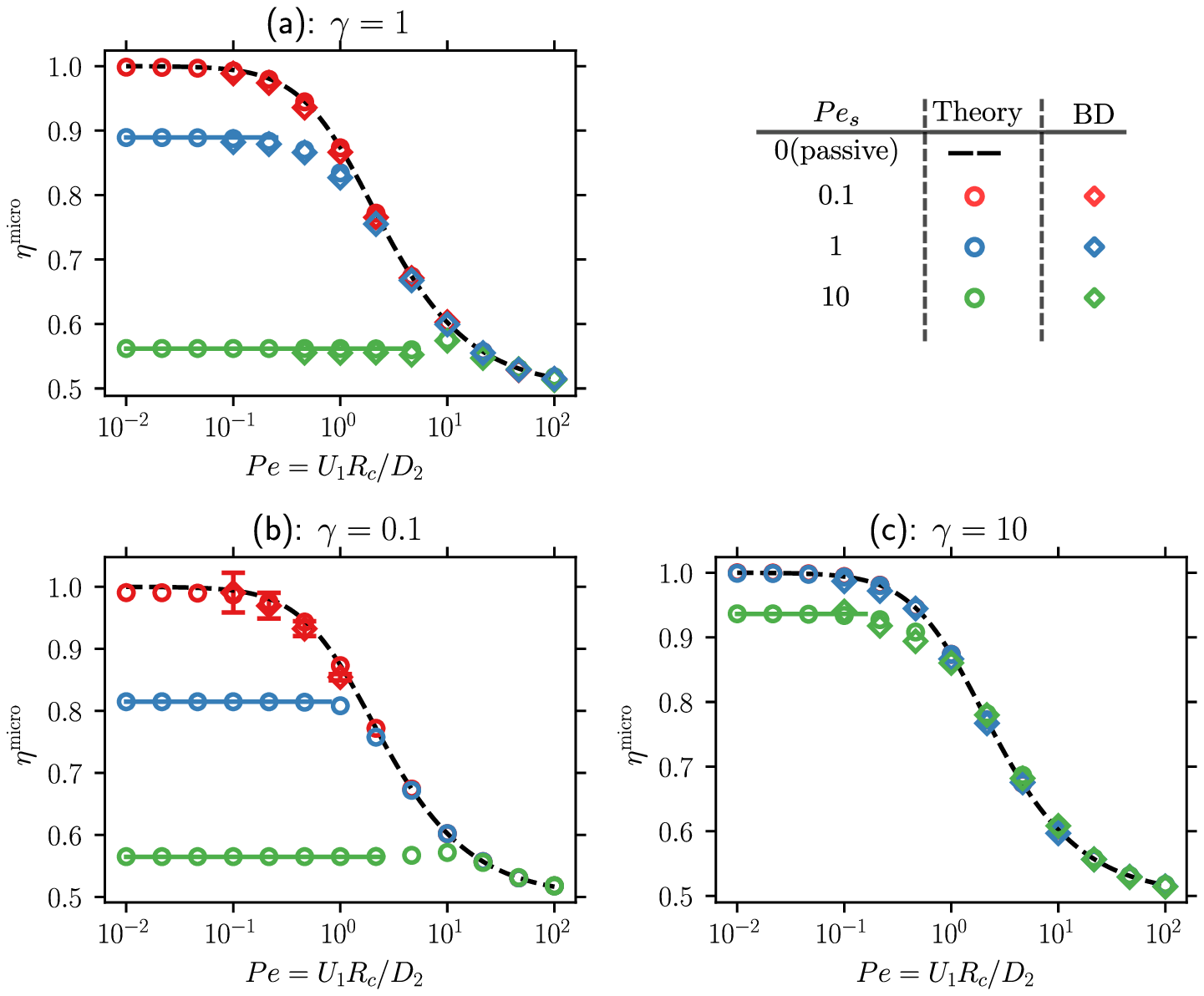}
  \caption{\label{fig:eta_micro_all}  The microviscosity of ABPs as a function of $Pe=U_1R_c/D_2$ for several values of $Pe_s=U_2R_c/D_2$ and $\gamma=R_c/\delta$. The dashed line denotes the results for passive Brownian suspensions ($Pe_s=0$). Circles are results from the numerical solutions of the Smoluchowski equation and diamonds are obtained from BD. The horizontal solid lines are the zero-velocity microviscosity $\eta_0^\text{micro}$ obtained in Sec. \ref{sec:slow-probe}. The values of $\gamma$ are fixed in each panel and are given by (a) $\gamma=1$, (b) $\gamma=0.1$, and (c) $\gamma=10$. Note that in panel (c), the results for $Pe_s=0.1$ and $Pe_s=1$ are visually indistinguishable. }
\end{figure*}

In Fig. \ref{fig:eta_micro_all}(a) we plot the microviscosity  $\eta^\text{micro}$ as a function of $Pe$ for several values of $Pe_s$ and $\gamma=1$. The solid line denotes the microviscosity of passive Brownian suspensions ($Pe_s=0$). Circles denote results from the numerical solution of the Smoluchowski equation and diamonds are obtained from BD. The solid horizontal lines denote the zero-velocity microviscosity discussed in Sec. \ref{sec:slow-probe}. For $Pe_s=0.1$, the ABPs are weakly active and the microviscosity approximates that of a passive suspension. As discussed in Sec. \ref{sec:slow-probe}, when the probe speed is small ($Pe \ll 1$), the ABPs exhibit swim-thinning in which the microviscosity decreases as $Pe_s$ increases. In the large $Pe$ limit, the activity of the ABPs is obscured by the rapid advection of the probe speed and, therefore, does not affect the microviscosity as $Pe\to \infty$. That is, regardless of $Pe_s$ (so long as it is finite), the microviscosity approaches that of the passive result of $1/2$ as $Pe \to \infty$.

For completeness, the variation of microviscosity for different values of $\gamma$ are presented in Figs. \ref{fig:eta_micro_all}(b) and \ref{fig:eta_micro_all}(c). The increase of $\gamma$ corresponds to the decrease in $\tau_R$ or an increase in the rotary diffusivity $D_R$. When $\gamma$ is large, e.g., $\gamma=10$ in Fig. \ref{fig:eta_micro_all}(c), the rotary diffusion is strong and the particles behave more like passive particles. Therefore, the swim-thinning is less prominent and the microviscosity is closer to that of passive suspensions. Conversely, for a small $\gamma$ as shown in \ref{fig:eta_micro_all}(b) for $\gamma=0.1$, the swim-thinning is stronger compared to the case shown in \ref{fig:eta_micro_all}(a) for $\gamma=1$.

\section{Concluding remarks}
\label{sec:cv-conclusion}
In this paper, we have considered the particle-tracking microrheology of an active colloidal suspension consisting of active Brownian spheres. The tracked particle, i.e., probe, is a passive colloidal sphere. When the probe is held fixed in an active suspension, the microstructure is isotropic but not in thermodynamic equilibrium. Because active particles self-propel, they accumulate at no-flux boundaries---a consequence of their out-of-equilibrium nature. In the context of microrheology when the probe is stationary, the number density at contact is higher than that in the bulk, far from the probe. Nevertheless, this isotropic state of active suspensions does not give rise to a net force on the probe, only an elevated osmotic pressure at contact compared to that in the bulk \citep{yan_brady_2015}. When the probe has a nonzero speed, the suspension microstructure is no longer isotropic, even for passive suspensions. Averaging the external force over Brownian fluctuations allows us to define a microviscosity similar to that in a passive suspension. By varying the prescribed speed of a CV probe, one can distort the suspension microstructure slightly ($Pe \ll 1$) or considerably ($Pe \gg 1$) from the isotropic steady state.

In the absence of HIs, the microrheological response of active suspensions originates from the interplay between the suspension activity and the excluded-volume interaction between the probe and the bath ABP. One manifestation of such a nontrivial interaction is the swim-thinning of the zero-velocity microviscosity. As the swim speed of the ABP increases, the zero-velocity microviscosity is lowered: $\eta_0^\text{micro} \to 1 $ as $Pe_s \to 0$ and $\eta_0^\text{micro} \to 1/2 $ as $Pe_s \to \infty$. This reduction in the suspension microviscosity below the passive result is a consequence of the out-of-equilibrium nature of ABPs. Our analysis shows that microrheology can be a useful tool to examine the mechanical response of other active or out-of-equilibrium materials.

In general, for finite activity, the suspension  exhibits a velocity-thinning behavior similar to that of passive suspensions but with a lowered $\eta_0^\text{micro}$. The high $Pe$ microviscosity of colloidal suspensions does not depend on activity due to the obscuring effect of the rapid advection of the probe. 

We note that both the swim-thinning and the convergence of the microviscosity of active suspensions to that of the passive result in the large $Pe$ limit have been observed by \citet{Burkholder2020}. In their work, an approximate solution to the Smoluchowski equation is considered using the closure $\bQ = \int (\bq\bq-\bI/d)gd \bq =\bm{0}.$ With $\bQ=\bm{0}$, one only needs to solve the equations governing the number density ($n=\int g d\bq $) and the polar order ($\polarorder = \int g \bq d\bq$) instead of the full probability $g$. For weak activity, the number density and the resulting microviscosity obtained using the $\bQ =\bm{0}$ closure agrees well with the full solution of the Smoluchowski equation. Generally for finite activity and finite $Pe$, the solution obtained using the $\bQ =\bm{0}$ closure does not agree well with the full solution of the Smoluchowski equation. In the present work, instead of a closure, we have solved the full Smoluchowski equation numerically and have shown that the results obtained from this continuum approach agree with those obtained from BD.

One interesting feature to note is that the microviscosity obtained in the successive limits $Pe \to 0$ and then $Pe_s \to \infty$ is the same as the limit $Pe \to \infty$ but with $Pe_s$ being finite (or zero). This can be understood by recalling the boundary layer structures in these two limits. For $Pe \to \infty$, there is an advective accumulation boundary layer in the front  and an empty wake devoid of particles in the back regardless of activity (provided that the swimming motion is not covarying with the probe speed, i.e., $Pe_s \ll Pe$ as $Pe \to \infty$). On the other hand, in the successive limits by first talking $Pe\to 0$ and then $Pe_s \to \infty$, there is again an accumulation boundary layer ( but due to swimming) for  orientations pointing into the probe. At contact, there are no particles with orientations pointing away from the probe in the limit $Pe_s \to \infty$. Structurally, these two limits share the behavior that only half of the domain has particles---half of the local orientation space for the first limit and half of the physical space for the second limit. The resulting net asymmetry of the density distribution at contact in both cases are identical and gives rise to the same microviscosity.

We expect our results to hold whenever the ``hard-sphere'' radii of the particles are much larger than the hydrodynamic radii so that HIs can be neglected. In a theoretical analysis, one can tune the strength of HIs using the excluded-annulus model \citep{brady1997microstructure,bergenholtz_brady_vicic_2002,khair_brady_2006}.  For passive colloids, HIs between particles give rise to a thickening behavior in the large $Pe$ limit \citep{khair_brady_2006,swan2013}. For large $Pe$,  we expect the microrheology of active particles to be indistinguishable from that of passive particles regardless of the inclusion of HIs.

Compared to passive colloids \citep{meyer2006,Sriram2009,Sriram2010}, experimental study of the forced microrheology of active particles are still lacking.  Notably, \citet{Ahmed22} studied the dynamics of a passive colloid in a suspension of swimming \textit{E. coli} using an optical trap; they observed a thinning behavior and quantified the force fluctuations on the immersed colloid. We hope that our theoretical analysis can prompt new experimental investigations in this area. 

To conclude, we have shown that in the absence of HIs the microviscosity of active suspensions are always positive. In particular, the swim-thinning in the low-$Pe$ limit at most can reduce the microviscosity by $1/2$, which is still positive. In a separate paper, we will show that a negative microviscosity can be obtained if some hydrodynamic effects are included. This suggests that the existence of a negative microviscosity is the result of HIs between the probe and the active bath particles.  In the present study, HIs are unaccounted for and the reduction in the microviscosity observed in the low-$Pe$ limit is due to activity.

\begin{acknowledgments}
    We thank Hyeongjoo Row for useful discussions. This work is supported by the National Science Foundation under Grant No. CBET 1803662.
\end{acknowledgments}

\section*{Author Declarations}
\subsection*{Conflict of Interest}
The authors have no conflicts to disclose.

\appendix

\section{Orientational moments for a slow probe}
\label{sec:CV-moments-appendix}

For a slow probe ($Pe \ll 1$), the suspension microstructure is expanded as $g = g_0(\br, \bq) +Pe g_1(\br, \bq)+\cdot\cdot\cdot$, where in any dimension we have at $O(1)$,
\begin{align}
    \label{eq:CV-g0-operator-form}
        &\nabla_r\cdot\left(Pe_s\bq g_0 - \nabla_r g_0 \right) -\gamma^2 \nabla_R^2 g_0=0,\\ 
    \label{eq:CV-g0-bc-operator-form}
        & \bn \cdot\left(Pe_s\bq g_0 - \nabla_r g_0 \right)=0\quad\text{at}\quad r=1,\\
        & g_0 \to \frac{1}{\Omega_d}\quad\text{as}\quad r\to \infty.
    \end{align}
    The $O(Pe)$ equation is nonhomogeneous, i.e., forced by the probe advection of the $g_0$ field, which reads
    \begin{align}
    \label{eq:CV-g1-operator-form}
        &\nabla_r\cdot\left(Pe_s\bq g_1 - \nabla_r g_1 \right) -\gamma^2 \nabla_R^2 g_1=\bUh_1\cdot\nabla_r g_0,\\ 
    \label{eq:CV-g1-bc-operator-form}
        & \bn \cdot\left(Pe_s\bq g_1 - \nabla_r g_1 \right)=\bn\cdot\bUh_1 g_0 \quad\text{at}\quad r=1,\\
        & g_1 \to 0\quad\text{as}\quad r\to \infty.
    \end{align}

Though in the main text the full Smoluchowski equation  (in 2D) is solved, it is useful to examine the symmetries of the orientational moments in the slow probe limit. The zeroth moment of Eq. \eqref{eq:CV-g0-operator-form} gives the governing equation for the $O(1)$ number density, which reads 
\begin{align}
\label{eq:CV-n0-eq}
    &\nabla_r\cdot\left( Pe_s \polarorder_0 -\nabla_r n_0\right)=0,\\
    &\bn \cdot\left( Pe_s \polarorder_0 -\nabla_r n_0\right)=0\quad\text{at}\quad r=1,\\
    & n_0 \to 1 \quad\text{as}\quad r \to \infty.
\end{align}
Here, $n_0 = \int g_0 d \bq $ is the number density and $\polarorder_0 = \int g_0 \bq d\bq$ is the polar order when the probe is fixed. The polar order satisfies 
\begin{align}
    &\nabla_r\cdot\left[ Pe_s \left( \bQ_0 + \frac{1}{d}n_0\bI \right) - \nabla_r\polarorder_0\right]+ (d-1)\gamma^2\polarorder_0=0,\\ 
    &\bn \cdot\left[ Pe_s \left( \bQ_0 + \frac{1}{d}n_0\bI \right) - \nabla_r\polarorder_0\right]=\bm{0}\quad\text{at}\quad r=1,
\end{align}
where  $\bQ_0 = \int (\bq\bq-\bI/d)g_0d\bq$ is the nematic field.

The spherical symmetry of the domain dictates that \citep{yan_brady_2015}
\begin{align}
\label{eq:CV-n0-symmetry}
    n_0(\br) &= p_0(r), ~\polarorder_0(\br) = \br p_1(r),\nonumber \\
    \bQ_0(\br)&=\left(\br\br-\frac{1}{d}r^2\bI \right)p_2(r),
\end{align}
where $p_0$--$p_2$ are scalar functions of the radial coordinate. One cannot solve Eq. \eqref{eq:CV-n0-eq} without knowledge of the polar order $\polarorder_0$. In fact, this hierarchy of orientational moments continue indefinitely. A truncation or closure is often used to close the set of moment equations. For example, the solutions to $n_0$ and $\polarorder_0$ are obtained by \citet{yan_brady_2015} with the closure $\bQ_0=\bm{0}$.

As a reference, we proceed to present the solution when $\bQ_0$ is included and a closure at the next order is used. The third orientational moment in 3D is 
\begin{equation}
    \tilde{\bB}_0 = \int \bq\bq\bq g_0 d\bq = \bB_0 + \boldsymbol{\alpha}\cdot\polarorder_0/5,
\end{equation}
where $\alpha_{ijkl}=\delta_{ij}\delta_{kl}+\delta_{il}\delta_{jk}+\delta_{ik}\delta_{jl}$ is an isotropic fourth order tensor. Assuming that $\bB_0=\bm{0}$, one can show that the general solutions are given by 
\begin{widetext}
\begin{align}
    p_0(r) &= C_0+ \sum_{k=1}^2 C_k\frac{1}{r} \exp\left[-\lambda_k (r-1)\right] +\frac{C_3}{r},\\
    p_1(r) &= - \sum_{k=1}^2 \frac{C_k}{Pe_s} \left(\frac{\lambda_k}{r^2}+\frac{1}{r^3}\right)\exp\left[-\lambda_k (r-1)\right]+ \frac{C_3 Pe_s}{6 \gamma ^2 r^3},\\
    p_2(r) &= \sum_{k=1}^2 C_k\frac{3 \lambda_k^2-6\gamma^2-Pe_s^2}{2Pe_s^2} \left(\frac{1}{r^3}+\frac{3}{\lambda_k r^4}+\frac{3}{\lambda_k^2 r^5}\right)\exp\left[-\lambda_k (r-1)\right]+\frac{C_3 Pe_s^2}{30 \gamma ^4 r^5},
\end{align}
\end{widetext}
where
\begin{align}
    \lambda_1 &= \frac{\sqrt{3 Pe_s^2+40 \gamma ^2+\sqrt{9 Pe_s^4+40 \gamma ^2 Pe_s^2+400 \gamma ^4}}}{\sqrt{10}},\\
    \lambda_2 &= \frac{\sqrt{3 Pe_s^2+40 \gamma ^2-\sqrt{9 Pe_s^4+40 \gamma ^2 Pe_s^2+400 \gamma ^4}}}{\sqrt{10}}.
\end{align}
The no-flux condition of $n_0$ means that $C_3=0$ and the far-field condition gives $C_0=1$. The other two integration constants $C_1$ and $C_2$ can be obtained from the no-flux conditions for $\polarorder_0$ and $\bQ_0$.

It is worthwhile to compare the results obtained with $\bQ_0=\bm{0}$ and $\bB_0=\bm{0}$. In particular, for $Pe_s \gg 1$ and $\gamma=O(1)$, the $\bQ_0$ closure predicts the correct scaling of the number density at contact [$n_c =O(Pe_s)$] while the $\bB_0$ closure gives a finite density. This comparison implies that a higher order closure is  not necessarily more accurate.

The zeroth moment of $g_1$ satisfies 
\begin{align}
        &\nabla_r\cdot\left( Pe_s \polarorder_1 -\nabla_r n_1\right)=\bUh_1\cdot\nabla_r n_0,\\
    &\bn \cdot\left( Pe_s \polarorder_1 -\nabla_r n_1\right)=\bn\cdot\bUh_1 n_0\quad\text{at}\quad r=1,\\
    & n_1 \to 0 \quad\text{as}\quad r \to \infty,
\end{align}
where $n_1$ and $\polarorder_1$ are similarly defined but for $g_1$; they are the leading-order disturbances to $n_0$ and $\polarorder_0$, respectively,  due to the weak probe motion. With \eqref{eq:CV-n0-symmetry}, it is straightforward to see that the $O(Pe)$ moments satisfy 
\begin{align}
    n_1(\br) &= \bUh_1\cdot\br p_3(r),\\ 
    \polarorder_1(\br) &= \bUh_1 p_4(r) + \bUh_1\cdot\br\br p_5(r),
\end{align}
where $p_3$-$p_5$ are unknown scalar radial functions. This is a manifestation of the so-called linear response  in which the disturbance fields are proportional to the vector $\bU_1$---the weak driving force.

In 2D, from \eqref{eq:eta-micro-2d}, we have $\eta_0^\text{micro} = p_3(1)$. Similarly in 3D, \eqref{eq:eta-micro-3d} gives $\eta_0^\text{micro} = 2 p_3(1)$. In other words, the zero-velocity microviscosity is determined from the contact value of $p_3$.

\section{The slow-swimming limit}
\label{sec:slow-swimming}
In the slow-swimming limit, characterized by $Pe_s \ll 1$, the probability distribution of bath colloids can be expanded as 
\begin{align}
    g(\br, \bq) = g_0(\br, \bq) +Pe_s g_1(\br, \bq) + Pe_s^2 g_2(\br, \bq) +\cdot\cdot\cdot.
\end{align}
Inserting the series into \eqref{eq:CV-smol-g}, \eqref{eq:CV-no-flux-g}, and \eqref{eq:CV-far-field-g}, we can solve the problem order by order.

At $O(1)$, the bath colloids are not self-propelling and the governing equations in the vector form are 
 \begin{align}
     &Pe \bUh_1\cdot \nabla_r n_0 + \nabla_r^2 n_0 = 0,\\
     & Pe \bn\cdot\bUh_1 n_0 + \bn\cdot\nabla_r n_0 =0\quad\text{at}\quad r=1,\\
     & n_0 \to 1\quad \text{as}\quad r \to \infty,
 \end{align}
 where $n_0(\br) = \int g_0 d\bq $ and $g_0 = n_0/\Omega_d$. Clearly, the equations at $O(1)$ govern the CV microrheology of passive colloids \citep{SquiresBrady2005}.
 
The problem at $O(Pe_s^k)$ ($k=1,2,...$) satisfies 
\begin{align}
    &\nabla_r\cdot\left(Pe \bUh_1 g_k + \nabla_r g_k  \right) +\gamma^2 \nabla_R^2 g_1 = \bq\cdot\nabla_r g_{k-1},\\ 
    & \bn\cdot\left(Pe \bUh_1 g_k + \nabla_r g_k \right)=\bn\cdot\bq g_{k-1}\quad\text{at}\quad r=1,\\
    & g_k \to 0 \quad\text{as}\quad r\to \infty.
\end{align}
 One can see that the solution at $O(Pe_s)$ has the structure $g_1(\br, \bq) = \bq\cdot\bG_1(\br)$, where $\bG_1(\br)$ is a vector-valued function of $\br$. This means that the number density at $O(Pe_s)$ vanishes, $n_1 = \int g_1 d\bq \equiv 0$. We note that the polar order ($\polarorder_1 = \int \bq g_1 d\bq$) is nonzero and is responsible for driving a density distribution at the next order [$O(Pe_s^2)$]. This structure ultimately leads to the fact that 
 \begin{align}
     \eta^\text{micro}(Pe, Pe_s, \gamma) = \eta^\text{micro}_\text{passive} + Pe_s^2 \eta_2^\text{micro} +\cdot\cdot\cdot.
 \end{align}
 In other words, the swim-thinning discussed in the main text occurs at $O(Pe_s^2)$.

\bibliography{reference}

\end{document}